\documentclass[twocolumn,pra,showpacs,showkeys,floatfix,amsmath,amssymb,superscriptaddress]{revtex4}
\usepackage{graphicx,exscale,bm,color}

\begin{document}
\title{Task-dependent control of open quantum systems} \date{\today}
\author{Jens Clausen}
\affiliation{Institute for Quantum Optics and Quantum Information,
  Austrian Academy of Sciences, Technikerstr. 21a, A-6020 Innsbruck}
\affiliation{Institute for Theoretical Physics, University of
  Innsbruck, Technikerstr. 25, A-6020 Innsbruck, Austria}
\author{Guy Bensky}
\author{Gershon Kurizki}
\affiliation{Department of Chemical Physics,
             Weizmann Institute of Science,
             Rehovot,
             76100, Israel}
\begin{abstract}
  We develop a general optimization strategy for performing a chosen
  unitary or non-unitary task on an open quantum system. The goal is
  to design a controlled time-dependent system Hamiltonian by
  variationally minimizing or maximizing a chosen function of the
  system state, which quantifies the task success (score), such as
  fidelity, purity, or entanglement.  If the time-dependence of the
  system Hamiltonian is fast enough to be comparable or shorter than
  the response-time of the bath, then the resulting non-Markovian
  dynamics is shown to optimize the chosen task score to second order
  in the coupling to the bath.  This strategy can protect a desired
  unitary system evolution from bath-induced decoherence, but ca also
  take advantage of the system-bath coupling so as to realize a
  desired non-unitary effect on the system.
\end{abstract}
\maketitle
\section{Introduction}
\label{sec1}
Due to the ongoing trends of device miniaturization, increasing
demands on speed and security of data processing, along with
requirements on measurement precision in fundamental research, quantum
phenomena are expected to play an increasing role in future
technologies. Special attention must hence be paid to omnipresent
decoherence effects. These may have different physical origins, such
as coupling of the system to an external environment (bath)
\cite{bookBreuer} or to internal degrees of freedom of a structured
particle \cite{BarGil06}, noise in the classical fields controlling
the system, or population leakage out of a relevant system subspace
\cite{WKB09}.  Formally, their consequence is always a deviation of
the quantum state evolution (error) with respect to the expected
unitary evolution if these effects are absent \cite{krausNielsen}.  In
operational tasks such as the preparation, transformation,
transmission, and detection of quantum states, these environmental
couplings effects are detrimental and must be suppressed by strategies
known as dynamical decoupling \cite{vio99,uhr07,kuo11,cai11}, or the more
general dynamical control by modulation
\cite{kof01,Barone04,kof04,JPB,gkl08}. 

There are however tasks which cannot be implemented by unitary
evolution, in particular those involving a change of the system's
state entropy
\cite{Alicki79,Lindbladbook}.
Such tasks necessitate a coupling to a bath and their efficient
implementation hence requires {\em enhancement} of this coupling.
Examples are the use of measurements to cool (purify) a system
\cite{gor09,gorennjp12,Gonzalo10,cha11} or manipulate its state
\cite{Harel96,kof00,Opatrny00} or harvest and convert energy from the
environment \cite{Scully03,gro10,scho11,sch04,bla11}.

A general task may also require state and energy transfer
\cite{esc11}, or entanglement \cite{vol11} of non-interacting parties
via {\em shared} modes of the bath \cite{durga11,gor06b}
which call for maximizing the shared (two-partite) couplings with the
bath, but suppressing the single-partite couplings.

It is therefore desirable to have a general framework for optimizing
the way a system interacts with its environment to achieve a desired
task.  This optimization consists in adjusting a given ``score'' that
quantifies the success of the task, such as the targeted fidelity,
purity, entropy, entanglement, or energy by dynamical modification of
the system-bath coupling spectrum on demand. The goal of this work is
to develop such a framework.

The remainder of the paper is organized as follows. In Sec.~\ref{sec2}
we state the problem formally and provide general expressions for the
change of the score over a fixed time interval as operator and matrix
spectral overlap.  In Sec.~\ref{sec3} we discuss a general solution in
terms of an Euler-Lagrange optimization. In Sec.~\ref{sec4} we apply
the approach to the protection of a general quantum gate, which
requires minimizing any coupling to the bath, whereas in
Sec.~\ref{sec5} we consider the complementary case of {\em enhancing}
the system-bath coupling in order to modify the purity (entropy) of a
qubit.  Open problems are outlined and an outlook is presented in
Sec.~\ref{sec6}. Supplementary information is provided in
Apps.~\ref{secA1} and \ref{secA2}.
\section{Overlap-integral formalism}
\label{sec2}
\subsection{Fixed time approach}
Assume that a quantity of interest (``score'') can be written as a
real-valued function $P(t)$ $\!=$ $\!P[\hat{\varrho}(t)]$ of the
system state $\hat{\varrho}(t)$ at a given time $t$.  This might be,
for example, a measure of performance of some input-output-device that
is supposed to operate within a predefined cycle/gate time $t$.
Depending on the physical problem and model chosen, extensions and
generalizations are conceivable, such as a comparison of the outcome
for different $t$ (on a time scale set by a constraint) \cite{esc11},
a time average $P$ $\!=$
$\!\int\mathrm{d}\tau{f}(\tau)P[\hat{\varrho}(\tau)]$ with some
probability density $f(\tau)$ \cite{cai10}, or a maximum $P(t)$ $\!=$
$\!\mathrm{max}_{\tau\in[0,t]}P[\hat{\varrho}(\tau)]$ \cite{sch11}.
Here we restrict ourselves to the ``fixed-time'' definition as given
above.  Our goal is to generate, by means of classical control fields
applied to the system, a time dependence of the system Hamiltonian
within the interval $0\le\tau\le{t}$ that adjusts $P(t)$ to a desired
value. In particular, this can be an optimum (i.e. maximum or minimum)
of the possible values of $P$. Assume that the initial system state
$\hat{\varrho}(0)$ is given. The {\em change} to the ``score'' over
time $t$, is then given by the first-order Taylor expansion in a
chosen basis
\begin{equation}
\label{dpoverl}
  P(t)
  \approx\sum_{mn}\frac{\partial{P}}{\partial\varrho_{mn}}\Delta\varrho_{mn}
  =\mathrm{Tr}\bigl(\hat{P}\Delta\hat{\varrho}\bigr),
\end{equation}
where the expansion coefficients
\begin{equation}
\label{defP}
  \left(\frac{\partial{P}}{\partial\varrho_{mn}}\right)_{t=0}
  \equiv(\hat{P})_{nm}
\end{equation}
are the matrix elements (in the chosen basis) of a Hermitian operator $\hat{P}$,
which is the gradient of $P[\hat{\varrho}(t)]$ with respect to $\hat{\varrho}$
at $t$ $\!=$ $\!0$, i.e., we may formally write
$\hat{P}$ $\!=$ $\!(\nabla_{\hat{\varrho}}P)_{t=0}^T$ $\!=$
$\!(\partial{P}/\partial\hat{\varrho})_{t=0}^T$.
In what follows, it is $\hat{P}$ which contains all information on the score
variable. Note that the transposition applied in Eq.~(\ref{defP}) simply allows to
express the sum over the Hadamard (i.e. entrywise) matrix product in
Eq.~(\ref{dpoverl}) as a trace of the respective operator product
$\hat{P}\Delta\hat{\varrho}$.

Let us illustrate this in two examples. If $P$ is the expectation
value of an observable (i.e. Hermitian operator) $\hat{Q}$, so that
$P$ $\!=$ $\!\mathrm{Tr}(\hat{\varrho}\hat{Q})$, then Eq.~(\ref{defP})
just reduces to this observable, $\hat{P}$ $\!=$ $\!\hat{Q}$.  If $P$
is the state purity, $P$ $\!=$ $\!\mathrm{Tr}(\hat{\varrho}^2)$, then
Eq.~(\ref{defP}) becomes proportional to the state, $\hat{P}$ $\!=$
$\!2\hat{\varrho}(0)$.  Note that the score $P$ is supposed to reflect
the environment (bath) effects and not the internal system dynamics.

Equation~(\ref{dpoverl}) implies that $\Delta\hat{\varrho}$ and with it
$P$ are small. Hence $\Delta\hat{\varrho}$ must refer to the
interaction picture and a weak interaction, while
$P[\hat{\varrho}(t)]$ should not be affected by the internal dynamics
[so that no separate time dependence emerges in Eq.~(\ref{dpoverl}), which
is not included in the chain-rule derivative]. In the examples above,
this is obvious for state purity, whereas an observable $\hat{Q}$
might be thought of co-evolving with the internal dynamics. 

Our starting point for what follows is simply the relation $P$
$\!=$ $\!\mathrm{Tr}(\hat{P}\Delta\hat{\varrho})$ with some Hermitian
$\hat{P}$, whose origin is not relevant.
\subsection{Role of averaged interaction energy}
Equation~(\ref{dpoverl}) expresses the score $P$ as
an overlap between the gradient $\hat{P}$ and the change of system
state $\Delta\hat{\varrho}$. In order to find expressions for
$P$ in terms of physically insightful quantities, we decompose
the total Hamiltonian into system, bath, and interaction parts,
\begin{equation}
  \hat{H}(t)=\hat{H}_{\mathrm{S}}(t)+\hat{H}_{\mathrm{B}}+\hat{H}_{\mathrm{I}},
\end{equation}
and consider the von Neumann equation of the total (system and environment)
state in the interaction picture,
\begin{equation}
\label{vNeq}
  \frac{\partial}{\partial{t}}\hat{\varrho}_{\mathrm{tot}}(t)
  =-\mathrm{i}[\hat{H}_{\mathrm{I}}(t),\hat{\varrho}_{\mathrm{tot}}(t)],
\end{equation}
[$\hat{H}_{\mathrm{I}}(t)$ $\!=$
$\!\hat{U}_{\mathrm{F}}^\dagger(t)\hat{H}_{\mathrm{I}}^{\mathrm{(S)}}(t)
\hat{U}_{\mathrm{F}}(t)$, (S) denoting the Schr\"odinger picture and
$\hat{U}_{\mathrm{F}}(t)$ $\!=$
$\!\mathrm{T}_+\mathrm{e}^{-\mathrm{i}\int_{0}^{t}\!
\mathrm{d}\tau[\hat{H}_{\mathrm{S}}(\tau)+\hat{H}_{\mathrm{B}}]}$].
Its solution can be written as Dyson (state) expansion
\begin{eqnarray}
  &&\hspace{-0.5cm}\hat{\varrho}_{\mathrm{tot}}(t)
  =\hat{\varrho}_{\mathrm{tot}}(0)
  +(-\mathrm{i})\int_{0}^{t}\mathrm{d}t_1
  [\hat{H}_{\mathrm{I}}(t_1),\hat{\varrho}_{\mathrm{tot}}(0)]
  \nonumber\\
  &&\hspace{-0.5cm}+(-\mathrm{i})^2
  \int_{0}^{t}\mathrm{d}t_1\int_{0}^{t_1}\mathrm{d}t_2
  [\hat{H}_{\mathrm{I}}(t_1),
  [\hat{H}_{\mathrm{I}}(t_2),\hat{\varrho}_{\mathrm{tot}}(0)]]+\ldots,\quad
\label{dyson}
\end{eqnarray}
which can be obtained either by an iterated integration of (\ref{vNeq}) or from
its formal solution
\begin{equation}
\label{fsol}
  \hat{\varrho}_{\mathrm{tot}}(t)=\hat{U}_{\mathrm{I}}(t)
  \hat{\varrho}_{\mathrm{tot}}(0)\hat{U}_{\mathrm{I}}^\dagger(t),
  \quad
  \hat{U}_{\mathrm{I}}(t)
  =\mathrm{T}_+\mathrm{e}^{-\mathrm{i}\int_{0}^{t}\!\mathrm{d}t^\prime
  \hat{H}_{\mathrm{I}}(t^\prime)},
\end{equation}
by applying the Magnus (operator) expansion
\begin{eqnarray}
  \hat{U}_{\mathrm{I}}(t)
  &=&\mathrm{e}^{-\mathrm{i}t\hat{H}_{\mathrm{eff}}(t)},
  \\
  \hat{H}_{\mathrm{eff}}(t)
  &=&\frac{1}{t}\int_{0}^{t}\mathrm{d}t_1\,\hat{H}_{\mathrm{I}}(t_1)
  \nonumber\\
  &&-\frac{\mathrm{i}}{2t}\int_{0}^{t}\mathrm{d}t_1
  \int_{0}^{t_1}\mathrm{d}t_2
  [\hat{H}_{\mathrm{I}}(t_1),\hat{H}_{\mathrm{I}}(t_2)]+\ldots,\quad\quad
\label{magnus}
\end{eqnarray}
expanding in (\ref{fsol}) the exponential $\hat{U}_{\mathrm{I}}$ $\!=$
$\!\hat{U}_{\mathrm{I}}(\hat{H}_{\mathrm{eff}})$ and sorting the terms
according to their order in $\hat{H}_{\mathrm{I}}$.

We assume that initially, the system is brought in contact with its
environment (rather than being in equilibrium with it), which
corresponds to factorizing initial conditions
$\hat{\varrho}_{\mathrm{tot}}(0)$ $\!=$
$\!\hat{\varrho}(0)\otimes\hat{\varrho}_{\mathrm{B}}$. The environment
is in a steady state $\hat{\varrho}_{\mathrm{B}}$,
$[\hat{\varrho}_{\mathrm{B}},\hat{H}_{\mathrm{B}}]$ $\!=$ $\!0$, so it
is more adequate to speak of a ``bath''.  Tracing over the bath in
Eq.~(\ref{dyson}) then gives the change of system state
$\Delta\hat{\varrho}$ $\!=$ $\!\hat{\varrho}(t)$ $\!-$
$\!\hat{\varrho}(0)$ over time $t$, which we must insert into
Eq.~(\ref{dpoverl}). We further assume a vanishing bath expectation
value of the interaction Hamiltonian,
\begin{equation}
\label{symmcond}
  \langle\hat{H}_{\mathrm{I}}\rangle_{\mathrm{B}}
  \equiv\mathrm{Tr}_{\mathrm{B}}(\hat{\varrho}_{\mathrm{B}}\hat{H}_{\mathrm{I}})
  =\hat{0}.
\end{equation}
As a consequence, the ``drift'' term corresponding to the first order
in Eq.~(\ref{dyson}) vanishes, and we only consider the second order
term as the lowest non-vanishing order approximation.  Finally, we
assume that the initial system state commutes with $\hat{P}$,
\begin{equation}
\label{condition}
  \bigl[\hat{\varrho}(0),\hat{P}\bigr]=0.
\end{equation}
In the language of control theory,
$\mathrm{Tr}[\hat{\varrho}(0)\hat{P}]$ is a ``kinematic critical
point'' \cite{pec11} if Eq.~(\ref{condition}) holds, since
$\mathrm{Tr}[\mathrm{e}^{\mathrm{i}\hat{H}}\hat{\varrho}(0)
\mathrm{e}^{-\mathrm{i}\hat{H}}\hat{P}]$ $\!=$
$\!\mathrm{Tr}[\hat{\varrho}(0)\hat{P}]$ $\!+$
$\!\mathrm{i}\mathrm{Tr}(\hat{H}[\hat{\varrho}(0),\hat{P}])$ $\!+$
$\!\mathcal{O}(\hat{H}^2)$ for a small arbitrary system Hamiltonian
$\hat{H}$.  Since we consider $\hat{\varrho}$ in the interaction
picture, Eq.~(\ref{condition}) means that the score is insensitive (in
first order) to a bath-induced unitary evolution (i.e., a generalized
Lamb shift) \cite{durga11}.  The purpose of this assumption is only to
simplify the expressions, but it is not essential. Physically, one
may think of a fast auxiliary unitary transformation that is applied
initially in order to diagonalize the initial state in the eigenbasis
of $\hat{P}$. Modifications to be made if Eq.~(\ref{condition})
does not hold are provided in App.~\ref{secA2}.

\begin{figure}[ht]
\includegraphics[width=6cm]{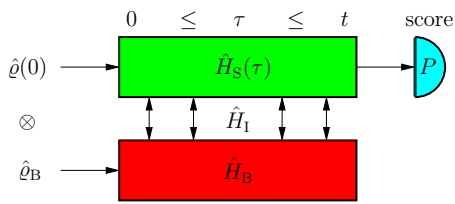}
\caption{\label{fig1}
Our control scheme: a system is brought in contact with a bath over a fixed time
interval $0\le\tau\le{t}$ during which the time dependence of the system
Hamiltonian $\hat{H}_{\mathrm{S}}(\tau)$ is chosen such that a given system
variable $P$ is adjusted to a desired value at the final time $t$.
}
\end{figure}

To lowest (i.e., second) order we then evaluate Eq.~(\ref{dpoverl})
for the score change as
\begin{equation}
\label{deltaP}
  P=t^2\bigl\langle[\hat{H},\hat{P}]\hat{H}\bigr\rangle,\quad
  \hat{H}=\frac{1}{t}\int_{0}^{t}\mathrm{d}\tau\,\hat{H}_{\mathrm{I}}(\tau),
\end{equation}
where $\langle\cdot\rangle$ $\!=$
$\!\mathrm{Tr}[\hat{\varrho}_{\mathrm{tot}}(0)(\cdot)]$.
This expresses the change of score in terms of the interaction Hamiltonian,
averaged in the interaction picture over the time interval of interest.
Our scheme is summarized in Fig.~\ref{fig1}.
\subsection{Spectral overlap}
\label{sec2c}
Alternatively, Eq.~(\ref{deltaP}) can be written as an overlap of
system-and bath matrices, which allows a more direct physical
interpretation. To do so, we assume $d$ dimensional Hilbert-space, and
expand the interaction Hamiltonian as a sum of products of system and
bath operators,
\begin{equation}
\label{HIdec}
  \hat{H}_{\mathrm{I}}=\sum_{j=1}^{d^2-1}\hat{S}_j\otimes\hat{B}_j,
\end{equation}
in such a way that $\langle\hat{B}_j\rangle$ $\!=$ $\!0$, which
ensures that Eq.~(\ref{symmcond}) is satisfied (otherwise we may shift
$\hat{B}_j^\prime$ $\!=$ $\!\hat{B}_j-\langle\hat{B}_j\rangle\hat{I}$,
$\hat{H}_{\mathrm{S}}^\prime$ $\!=$
$\!\hat{H}_{\mathrm{S}}+\sum_j\langle\hat{B}_j\rangle\hat{S}_j$).
Considering Eq.~(\ref{HIdec}) in the interaction picture and expanding
$\hat{S}_j(t)$ $\!=$ $\!\sum_k{\epsilon}_{jk}(t)\hat{S}_k$ in terms of
[Hermitian, traceless, orthonormalized to
$\mathrm{Tr}(\hat{S}_j\hat{S}_k)$ $\!=$ $\!d\,\delta_{jk}$] basis
operators $\hat{S}_j$, defines a (real orthogonal) rotation matrix
${\bm{\epsilon}}(t)$ in the system's Hilbert space, with elements
\begin{equation}
  {\epsilon}_{jk}(t)
  =\bigl\langle\hat{S}_j(t)\hat{S}_k\bigr\rangle_{\mathrm{id}},
\end{equation}
where $\langle\cdot\rangle_{\mathrm{id}}$ $\!=$
$\!\mathrm{Tr}[d^{-1}\hat{I}(\cdot)]$.  These elements of the matrix
${\bm{\epsilon}}(t)$ may be regarded as the dynamical correlation
functions of the basis operators.  Analogously, we define a bath
correlation matrix ${\bm{\Phi}}(t)$ with elements
\begin{equation}
  {\Phi}_{jk}(t)=\bigl\langle\hat{B}_j(t)\hat{B}_k\bigr\rangle_{\mathrm{B}}.
\end{equation}
It contains the entire description of the bath behavior in our
approximation.  Finally, we define a Hermitian matrix ${\bm{\Gamma}}$
with elements
\begin{equation}
\label{Gm}
  \Gamma_{kj}=\langle[\hat{S}_j,\hat{P}]\hat{S}_k\rangle,
\end{equation}
where $\langle\cdot\rangle$ $\!=$
$\!\mathrm{Tr}[\hat{\varrho}(0)(\cdot)]$. The matrix ${\bm{\Gamma}}$
may be understood as a representation of the gradient $\hat{P}$ with
respect to the chosen basis operators $\hat{S}_j$. Finally, we define
the bath and (finite-time) system spectra according to
\begin{eqnarray}
\label{Gdef}
  {\bm{G}}(\omega)
  &=&\int_{-\infty}^{\infty}\!\mathrm{d}t\;\mathrm{e}^{\mathrm{i}\omega{t}}\;
  {\bm{\Phi}}_{}(t),
  \\
  {\bm{\epsilon}_t}(\omega)
  &=&\frac{1}{\sqrt{2\pi}}\int_{0}^t\mathrm{d}\tau\,
  \mathrm{e}^{\mathrm{i}\omega\tau}{\bm{\epsilon}}(\tau).
\end{eqnarray}
This allows to express the score, Eq.~(\ref{deltaP}), as the matrix
overlap
\begin{eqnarray}
  {P}&=&\iint_{0}^{t}\mathrm{d}t_1\mathrm{d}t_2
  \mathrm{Tr}[{\bm{\epsilon}}^T(t_1){\bm{\Phi}}(t_1-t_2){\bm{\epsilon}}(t_2)
  {\bm{\Gamma}}]
\label{soverl1}
  \\
  &=&\int_{-\infty}^{\infty}\!\!\mathrm{d}\omega\,
  \mathrm{Tr}[{\bm{\epsilon}_t}^\dagger(\omega)
  {\bm{G}}(\omega){\bm{\epsilon}_t}(\omega){\bm{\Gamma}}]
\label{soverl2}
  \\
  &=&t\int_{-\infty}^{\infty}\!\!\mathrm{d}\omega\,
  \mathrm{Tr}[{\bm{F}_t}(\omega){\bm{G}}(\omega)].
\label{soverl3}
\end{eqnarray}
In Eq.~(\ref{soverl3}) we have used the cyclic property of the trace
to write the spectral overlap in a more compact form by combining the
rotation matrix spectra ${\bm{\epsilon}_t}(\omega)$ and the gradient
representation ${\bm{\Gamma}}$ to a system spectral matrix
\begin{equation}
  {\bm{F}_t}(\omega)=\frac{1}{t}
  {\bm{\epsilon}_t}(\omega){\bm{\Gamma}}{\bm{\epsilon}_t}^\dagger(\omega).
\end{equation}
Analogously, Eq.~(\ref{soverl1}) can be written in a more compact form
by introducing the matrix
\begin{equation}
\label{dm}
  {\bm{R}}(t_1,t_2)
  ={\bm{\epsilon}}^T(t_1){\bm{\Phi}}(t_1-t_2)
  {\bm{\epsilon}}(t_2).
\end{equation}

Equation~(\ref{soverl3}) is as a generalization of \citet{kof05} and
demonstrates that the change $P$ over a given time $t$ is determined
by the spectral overlap between system and bath dynamics, analogously
to DCM \cite{gkl08}, or the measurement induced quantum Zeno and
anti-Zeno control of open systems \cite{kof00,kof96}.  The
bath-spectral matrix ${\bm{G}}(\omega)$ must be positive semi-definite
for all $\omega$. If the same holds for the matrix
${\bm{F}_t}(\omega)$, then $P$ is always positive. Below we will
consider such a case where $P$ reflects a gate error and the goal is
then to minimize this error. The spectral overlap Eq.~(\ref{soverl3})
can be made as small as desired by a rapid modulation of the system,
such that the entire weight of the system spectrum is shifted beyond
that of the bath, which is assumed to vanish for sufficiently high
frequencies. Since this fast modulation may cause unbounded growth of
the system energy, a meaningful posing of the problem requires a
constraint.

In general, ${\bm{F}_t}(\omega)$ is Hermitian but need not necessarily
be positive semi-definite, depending on the choice of score as encoded
in ${\bm{\Gamma}}$. This reflects the fact that $P$ can increase or
decrease over $t$. Depending on the application, our goal can
therefore also be to maximize $P$ with positive and negative sign. In
what follows we will consider the question how to find a system
dynamics that optimizes the score.
\section{\label{sec3}
         Euler-Lagrange Optimization}
\subsection{Role of control, score and constraint}
Our considerations in the previous section suggest to define our control
problem in terms of a triple $(\bm{f},P,E)$ consisting of a control $\bm{f}$, a
score $P$, and a constraint $E$.

The \emph{control} is a set of real parameters $f_l$, which have been
combined to a vector $\bm{f}$. These can either be timings,
amplitudes, and/or phases of a given number of discrete pulses, or
describe a time-continuous modulation of the system.  Here, we focus
on time-dependent control, where the $f_l(\tau)$ parametrize the
system Hamiltonian as $\hat{H}_{\mathrm{S}}$ $\!=$
$\!\hat{H}_{\mathrm{S}}[{\bm{{f}}}(\tau)]$, or the unitary evolution
operator $\hat{U}(\tau)$ $\!=$ $\!T_+
\mathrm{e}^{-\mathrm{i}\int_0^\tau\mathrm{d}\tau^\prime\,\hat{H}_{\mathrm{S}}
  (\tau^\prime)}$ $\!\equiv$ $\!\hat{U}[{\bm{{f}}}(\tau)]$. A direct
parametrization of $\hat{U}$ avoids the need of time-ordered
integration of its exponent. The $\hat{U}(\tau)$ thus obtained
\cite{cla10} can be then used to calculate the system Hamiltonian
$\hat{H}_{\mathrm{S}}(\tau)$ $\!=$
$\!\mathrm{i}[\frac{\partial}{\partial{t}}
\hat{U}(\tau)]\hat{U}^\dagger(\tau)$.

Two explicit examples of the score $P$ pertain to the fidelity of
$\hat{P}$ with a given pure state $F_{\Psi}$ $\!=$
$\!\langle\Psi|\hat{\varrho}|\Psi\rangle$, (for which $\hat{P}$ $\!=$
$\!|\Psi\rangle\langle\Psi|$), or to the von Neumann entropy which we
can approximate (for nearly pure states) by the linear entropy, $S$
$\!=$ $\!-k\mathrm{Tr}(\hat{\varrho}\mathrm{ln}\hat{\varrho})$
$\!\approx$ $\!S_{\mathrm{L}}$ $\!=$
$\!k[1-\mathrm{Tr}(\hat{\varrho}^2)]$, [for which $\hat{P}$ $\!=$
$\!-2k\hat{\varrho}(0)$]. The latter score can be used to maximize the
fidelity with the maximally mixed state $\hat{\varrho}$ $\!\sim$
$\!\hat{I}$ (for which $S_{\mathrm{L}}$ becomes maximum), or to
maximize the concurrence $C_{|\Psi_{\mathrm{AB}}\rangle}$ $\!=$
$\!\sqrt{2(1-\mathrm{Tr}\hat{\varrho}_{\mathrm{A}}^2)}$,
$\hat{\varrho}_{\mathrm{A}}$ $\!=$ $\!\mathrm{Tr}_{\mathrm{B}}
|\Psi_{\mathrm{AB}}\rangle\langle\Psi_{\mathrm{AB}}|$, as a measure of
entanglement of a pure state $|\Psi_{\mathrm{AB}}\rangle$ of a
bipartite system.

If a \emph{constraint} is required to ensure the existence of a finite
(physical) solution, its choice should depend on the most critical
source of error. An example is the average speed with which the
controls change, $E$ $\!=$
$\!\int_{0}^{t}\mathrm{d}{\tau}\,\dot{{\bm{{f}}}}^2({\tau})$, which
depend on the control bandwidth in the spectral domain. A
parametrization-independent alternative is the mean square of the
modulation energy, $E_{}$ $\!=$ $\!\int_{0}^{t}\!\mathrm{d}{\tau}\,
\bigl\langle(\Delta\hat{H})^2({\tau}) \bigr\rangle_{\mathrm{id}}$,
where $\langle\cdot\rangle_{\mathrm{id}}$ refers to a maximally mixed
state and hence to a state-independent norm, and $\Delta\hat{H}$ is
the difference between the modulated and unmodulated (natural) system
Hamiltonians.
\subsection{A projected gradient search}
\begin{figure}[ht]
\includegraphics[width=4cm]{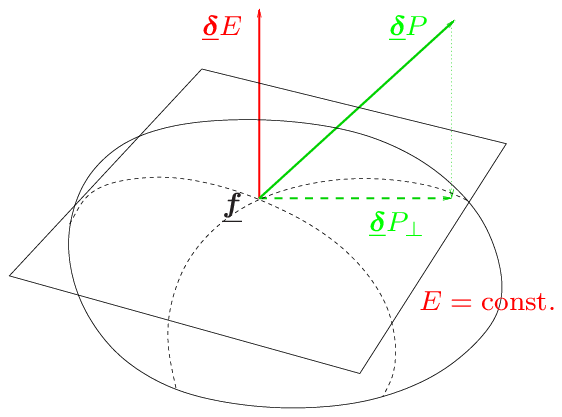}
\hspace{0.4cm}
\includegraphics[width=4cm]{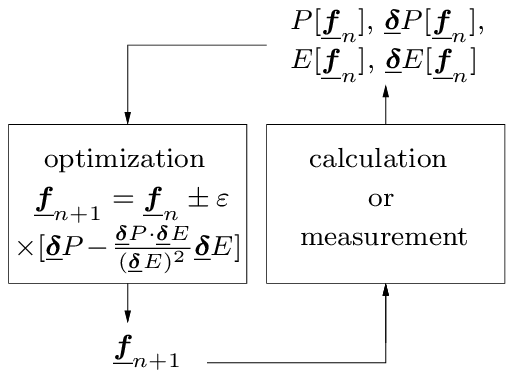}
\caption{\label{fig2}
Left: Optimization of the score $P({\bm{{f}}})$ subject to a constraint
$E({\bm{{f}}})$ in control space $\{{\bm{{f}}}\}$ by
walking along the component ${\bm{\delta}}{P}_\perp$ of the gradient
${\bm{\delta}}{P}$ orthogonal to ${\bm{\delta}}{E}$. Right:
Resulting iteration consisting in steps determined by a small parameter
$\epsilon$ which yields a local solution depending on the starting point
${\bm{{f}}}_0$.
}
\end{figure}

We want to find controls $\bm{f}$ that optimize a score $P(\bm{f})$
subject to a constraint $E(\bm{f})$. A numerical local optimization
can be visualized in parameter space as shown in Fig.~\ref{fig2}.

We start at some initial point $\bm{f}_0$ for which $E(\bm{f}_0)$ is
the desired value of the constraint. Simply following the gradient
$\bm{\delta}P$ would maximize or minimize $P$, but also change $E$. To
optimize $P$ while keeping $E$ constant, we therefore move along the
projection of $\bm{\delta}P$ orthogonal to $\bm{\delta}E$, i.e., along
$\bm{\delta}P_{\perp}$ $\!=$ $\!\bm{\delta}P$ $\!-$
$\!\frac{\bm{\delta}P\cdot\bm{\delta}E}{(\bm{\delta}E)^2}\bm{\delta}E$.
Since the gradients depend on $\bm{f}$, the iteration consists of
small steps $\bm{f}_{n+1}$ $\!=$ $\!\bm{f}_n$ $\!\pm$
$\!\epsilon\bm{\delta}P_{\perp}(\bm{f}_n)$, $\epsilon$ $\!\ll$ $\!1$.
Assuming that neither $\bm{\delta}P$ nor $\bm{\delta}E$ vanish, the
iteration will come to a halt where $\bm{\delta}P_{\perp}$ vanishes,
because the gradients are parallel,
\begin{equation}
\label{ELG}
  {\bm{\delta}}P
  =\lambda{\bm{\delta}}E.
\end{equation}
This condition constitutes the Euler-Lagrange (EL) equation of the
extremal problem, with the proportionality constant $\lambda$ being
the Lagrange multiplier. Its concrete form depends on the choice of
$P$ and $E$. Since the solutions of the EL optimization represent
local optima of the constrained $P$, we may repeat the search with
randomly chosen $\bm{f}_0$ a number of times and select the best
solution. The gradients at each point $\bm{f}_n$ may be obtained
either from a calculation based on prior knowledge of the bath or
experimentally from data measured in real time. A discretization of
the time interval $0$ $\!\le$ $\!\tau$ $\!\le$ $\!t$ then reduces the
variational $\bm{\delta}$ to a finite-dimensional vector gradient
$\bm{\nabla}$.
\section{Gate protection with BOMEC}
\label{sec4}
\subsection{Gate error as average fidelity decline}
A particular application of our formalism is decoherence protection of
a given quantum operation by bath-optimal minimal-energy control
(BOMEC) \cite{gkl08,cla10}. Consider the implementation of a
predetermined quantum gate, i.e., unitary operation within a given
``gate time'' $t$. It is sufficient to consider a pure input state
$|\Psi\rangle$.  In the interaction picture with respect to the
desired gate operation and in the absence of bath effects, we should
therefore observe at time $t$ the initial state $|\Psi\rangle$.  The
quantity of interest is here the fidelity
$\langle\Psi|\hat{\varrho}(t)|\Psi\rangle$, and we use the projector
$\hat{P}$ $\!=$ $\!\hat{\varrho}(0)$ $\!=$
$\!|\Psi\rangle\langle\Psi|$ as the gradient operator, so that
Eq.~(\ref{condition}) is satisfied and Eq.~(\ref{deltaP}) gives the
fidelity change as the score
\begin{equation}
  {P}=\langle\Psi|\Delta\hat{\varrho}|\Psi\rangle
  =-t^2\bigl\langle
  \langle\Psi|\hat{H}^2|\Psi\rangle-\langle\Psi|\hat{H}|\Psi\rangle^2
  \bigr\rangle_{\mathrm{B}},
\end{equation}
which is given by $\hat{H}$ defined in Eq.~(\ref{deltaP}). 

Since a quantum gate is supposed to act on an unknown input state, we
need to get rid of the dependence on $|\Psi\rangle$. One possibility
is to perform a uniform average over all $|\Psi\rangle$. We may apply
\begin{equation}
\label{dankert}
  \overline{\langle\Psi|\hat{A}|\Psi\rangle\langle\Psi|\hat{B}|\Psi\rangle}
  =\frac{\mathrm{Tr}\hat{A}\hat{B}+\mathrm{Tr}\hat{A}\mathrm{Tr}
  \hat{B}}{d(d+1)}
\end{equation}
\cite{dan05,sto05} which gives the average
\begin{equation}
\label{deltaPbar}
  \overline{{P}}
  =-t^2\frac{d}{d+1}
  \bigl\langle\hat{H}^2\bigr\rangle_{\mathrm{id}},
\end{equation}
where $\langle\cdot\rangle_{\mathrm{id}}$ $\!=$
$\!\mathrm{Tr}[d^{-1}\hat{I}\otimes\hat{\varrho}_{\mathrm{B}}(\cdot)]$.
In Eq.~(\ref{deltaPbar}) we have used
$\mathrm{Tr}_{\mathrm{S}}\hat{H}$ $\!=$ $\!\hat{0}$, which corresponds
to $\mathrm{Tr}\hat{S}_j$ $\!=$ $\!0$ in Sec.~\ref{sec2c}.  Because of
this and because of Eq.~(\ref{symmcond}),
$\langle\hat{H}\rangle_{\mathrm{B}}$ $\!=$
$\!\langle\hat{H}_{\mathrm{I}}\rangle_{\mathrm{B}}$ $\!=$ $\!\hat{0}$,
we have $\bigl\langle\hat{H}\bigr\rangle_{\mathrm{id}}$ $\!=$ $\!0$,
and Eq.~(\ref{deltaPbar}) also describes the variance
$\mathrm{Var}(\hat{H})$ $\!=$
$\!\bigl\langle\hat{H}^2\bigr\rangle_{\mathrm{id}}$ $\!-$
$\!\bigl\langle\hat{H}\bigr\rangle_{\mathrm{id}}^2$.  On the other
hand $\hat{P}$ $\!=$ $\!-2k\hat{\varrho}(0)$ that gives the change
$\Delta{S}$ of entropy $S$ $\!=$
$\!-k\mathrm{Tr}(\hat{\varrho}\mathrm{ln}\hat{\varrho})$ is (up to a
proportionality factor of $-2k$) the same as the $\hat{P}$ used here
to give the change of fidelity, we have $\Delta{S}$ $\!=$ $\!-2k{P}$.
If we define a \emph{gate error} $\mathcal{E}$ as the average fidelity
decline, $\mathcal{E}$ $\!=$ $\!-\overline{{P}}$, with
$\overline{{P}}$ given in Eq.~(\ref{deltaPbar}), we can summarize
the following proportionalities: gate error $\equiv$ average fidelity
decline $\sim$ average entropy increase (purity decline)
$\overline{\Delta{S}}$ $\sim$ square (variance) of the average
interaction energy $\hat{H}$:
\begin{equation}
\label{ge}
  \mathcal{E}\equiv-\overline{{P}}
  =\frac{\overline{\Delta{S}}}{2k}
  =t^2\frac{d}{d+1}
  \bigl\langle\hat{H}^2\bigr\rangle_{\mathrm{id}}
  =t^2\frac{d}{d+1}\mathrm{Var}(\hat{H}).
\end{equation}
In the matrix representation of Sec.~\ref{sec2c}, the average over the
initial states in the matrix ${\bm{\Gamma}}$ defined in
Eq.~(\ref{Gm}), gives $\overline{\bm{\Gamma}}$ $\!=$
$\!-\frac{d}{d+1}\bm{I}$ [using $\mathrm{Tr}(\hat{S}_j\hat{S}_k)$
$\!=$ $\!d\,\delta_{jk}$ and $\mathrm{Tr}\hat{S}_j$ $\!=$ $\!0$], so
that
\begin{equation}
  \mathcal{E}=\frac{d}{d+1}\int_{-\infty}^{\infty}\!\!\mathrm{d}\omega\,
  \mathrm{Tr}[\bm{\epsilon}_t(\omega)\bm{\epsilon}_t^\dagger(\omega)
  \bm{G}(\omega)]
\end{equation}
in agreement with \cite{cla10} [except a different normalization
$\mathrm{Tr}(\hat{S}_j\hat{S}_k)$ $\!=$ $\!2\delta_{jk}$ leading there to a
prefactor $\frac{2}{d+1}$].
From the requirement that $\mathcal{E}$ $\!\ge$ $\!0$ must hold for any positive
semi-definite matrix $\bm{\epsilon}_t(\omega)\bm{\epsilon}_t^\dagger(\omega)$,
we conclude that $\bm{G}(\omega)$ must be a positive semi-definite matrix for
any $\omega$. The task of BOMEC is then to find a system evolution
$\hat{U}(\tau)$ (cf. the control examples in the previous section) that
minimizes $\mathcal{E}$, subject to the boundary condition that the final
$\hat{U}(t)$ is the desired gate.
\subsection{Comparison of BOMEC with DD}
\begin{figure}[ht]
\includegraphics[width=4.2cm]{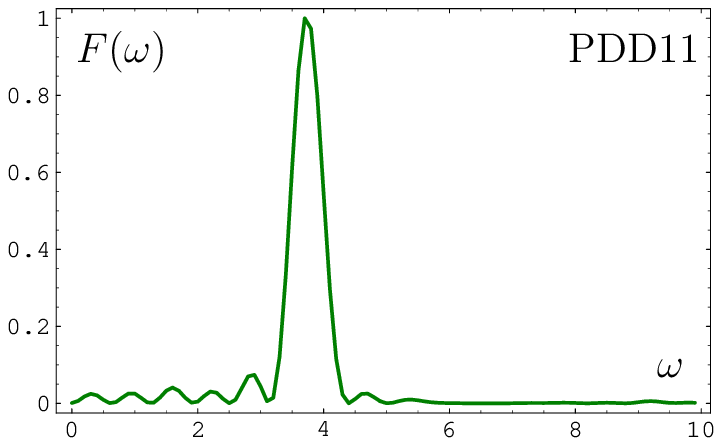}
\includegraphics[width=4.2cm]{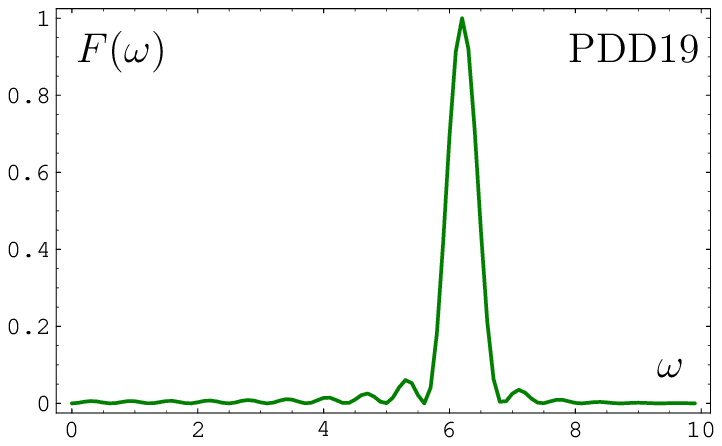}\\
\includegraphics[width=4.2cm]{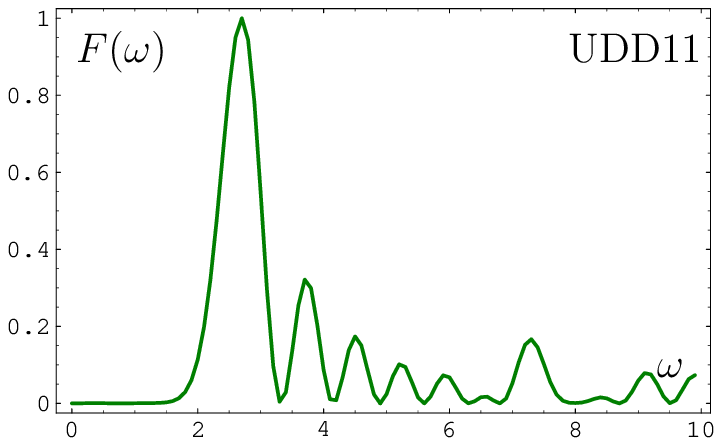}
\includegraphics[width=4.2cm]{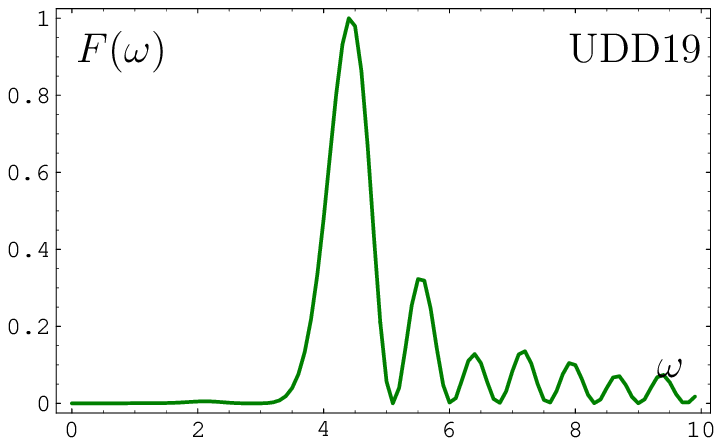}
\caption{\label{fig3}
System modulation spectra $F(\omega)$ generated by two methods of DD.
Upper row: periodic dynamical decoupling with $n$ $\pi$-pulses (PDDn),
lower row: Uhrig dynamical decoupling with the same number of $\pi$-pulses
(UDDn), compared for $n=11$ (left column) and $n=19$ (right column) pulses.
}
\end{figure}

It is appropriate to compare the effect of dynamical decoupling (DD)
\cite{vio99,uhr07} with that of BOMEC. DD does not change with the
bath spectrum $\bm{G}(\omega)$. With an increasing number of pulses,
DD shifts the weight of the system spectrum $\bm{F}(\omega)$ towards
higher frequencies, until the overlap Eq.~(\ref{soverl3}) has become
sufficiently small. This is illustrated for two different numbers of
pulses of periodic DD (PDD, pulses periodic in time), in the upper row
of Fig.~\ref{fig3} in the case of a 1D single qubit modulation (i.e.,
all pulses are given by an arbitrary but fixed Pauli matrix).

Aperiodic DD such as UDD \cite{uhr07} suppress low-frequency
components (to the left of the main peak) in the system spectrum,
which retain the system-bath coupling even if the main peak of the
system spectrum has been shifted beyond the bath cutoff frequency
(Fig.~\ref{fig3}). The plots indicate that this suppression of low
frequency components is achieved at the price of a smaller shift of
the main peak, i.e., shifting the main peak beyond a given cutoff
requires more pulses in UDD than in PDD.  Note that optimized DD
sequences with improved asymptotics exist \cite{kuo11}, which we will
not consider here.

\begin{figure}[ht]
\includegraphics[width=6cm]{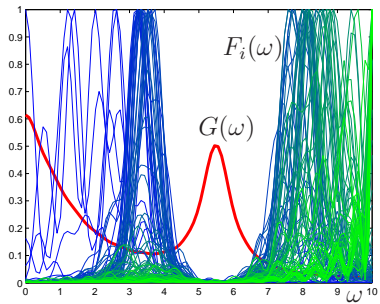}
\caption{\label{fig4}
BOMEC-minimization of the gate error for a single qubit $\pi$-gate caused by
pure dephasing with a given bath spectrum [$G(\omega)$, bold red line].
The $Z$-components of the obtained system modulation spectra $F_i(\omega)$ are
shown for energy constraints
$E_i$ $\!=$ $\!0.1+4(i-1)$, $i$ $\!=$ $\!1,2,\ldots,101$,
(thin lines, blue to green) individually scaled to 1.
}
\end{figure}

System modulation spectra obtained with BOMEC are shown in Fig.~\ref{fig4}.

The plot refers to a qubit subject to pure dephasing (i.e.,
$Z$-coupling) by a bath whose spectrum $G(\omega)$ has a Lorentzian
peak and low-frequency tail.  The BOMEC optimizes $\hat{U}(\tau)$
simultaneously for 3D Pauli matrix couplings to the bath (Z, Y and X).
The resulting system spectrum $F(\omega)$ is shown for different
energy constraints $E$ which are increased in small and equal steps.
For low $E$, $F(\omega)$ has a single peak on the left of the bath
peak. Increasing $E$ causes a second peak of $F(\omega)$ to emerge on
the right of the bath peak, which continues to grow, while the peak on
the left diminishes, until for high $E$, only the right peak remains.
Fig.~\ref{fig4} hence demonstrates that the spectrum $F(\omega)$
generated by BOMEC changes continuously as $E$ increases, but avoids
overlap with the maxima of $G(\omega)$ irrespective of $E$. BOMEC can
therefore be superior to all forms of DD including UDD, especially if
the bath has high cutoff but bandgaps at low frequencies.
\section{Purity control of a qubit}
\label{sec5}
To give an example of the opposite case, where the goal is to maximize
the system-bath coupling, we apply our approach of constrained
optimization to the linear entropy $S_{\mathrm{L}}$ $\!=$
$\!2[1-\mathrm{Tr}(\hat{\varrho}^2)]$ of a qubit. [Note that here
$S_{\mathrm{L}}$ has been normalized to 1 by setting the coefficient
$k$ $\!=$ $\!d/(d-1)$ $\!=$ $\!2$, cf.  Sec.~\ref{sec3}.] We assume an
initial mixture
\begin{equation}
\label{defp}
  \hat{\varrho}(0)=p|1\rangle\langle1|+(1-p)|0\rangle\langle0|
\end{equation}
of a ground (excited) state $|0\rangle$ ($|1\rangle$), where $0$
$\!\le$ $\!p$ $\!\le$ $\!0.5$ is related to $S_{\mathrm{L}}$ by
$p=(1-\sqrt{1-S_{\mathrm{L}}})/2$.  With $\hat{S}_j$ $\!=$
$\!\hat{\sigma}_j$ denoting for $d$ $\!=$ $\!2$ the Pauli matrices,
Eq.~(\ref{defp}) can be written in terms of $\hat{H}_0$ $\!=$
$\!\frac{\omega_0}{2}\hat{\sigma}_3$ as $\hat{\varrho}(0)$ $\!=$
$\!\frac{\mathrm{e}^{-\beta_{}\hat{H}_0}}
{\mathrm{Tr}(\mathrm{e}^{-\beta_{}\hat{H}_0})}$ $\!=$
$\!\frac{|1\rangle\langle1|}{1+\mathrm{e}^{\beta_{}\omega_0}}$ $\!+$
$\!\frac{|0\rangle\langle0|}{1+\mathrm{e}^{-\beta_{}\omega_0}}$, where
$\beta_{}$ $\!=$ $\!\frac{\ln(p^{-1}-1)}{\omega_0}$ is the inverse
temperature.  Purity and temperature are hence related via the energy
scale $\omega_0$.  Our goal is a constrained optimization of
$\Delta{S}_{\mathrm{L}}$, i.e., $\hat{P}$ $\!=$
$\!-4\,\hat{\varrho}(0)$ in Eq.~(\ref{defP}).  Unlike the gate error
Eq.~(\ref{ge}), $\Delta{S}_{\mathrm{L}}$ can be negative or positive,
which can be understood as cooling or heating, respectively.

\begin{figure}[ht]
\includegraphics[width=4.2cm]{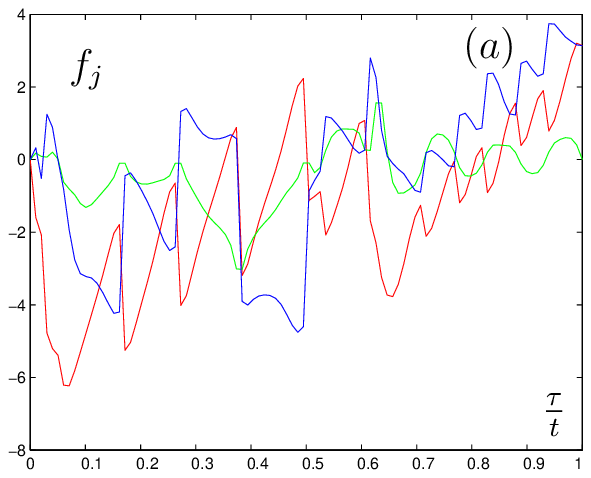}
\includegraphics[width=4.2cm]{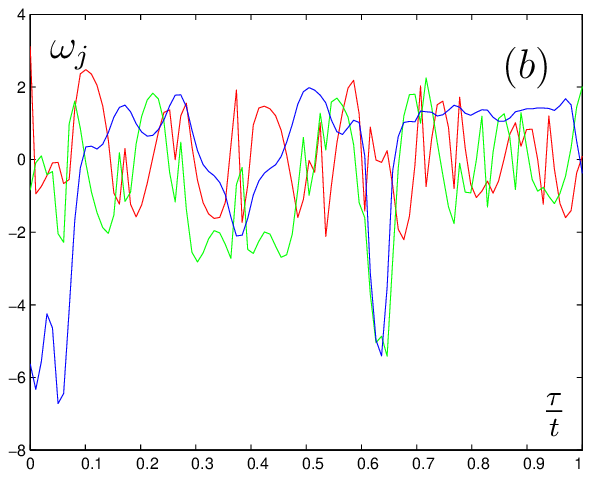}
\includegraphics[width=4.2cm]{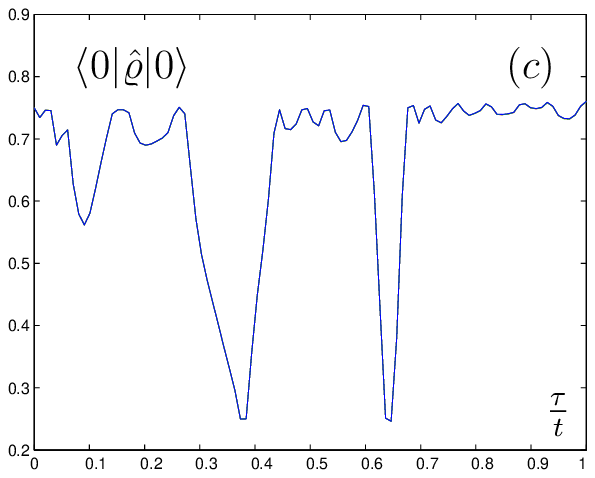}
\includegraphics[width=4.2cm]{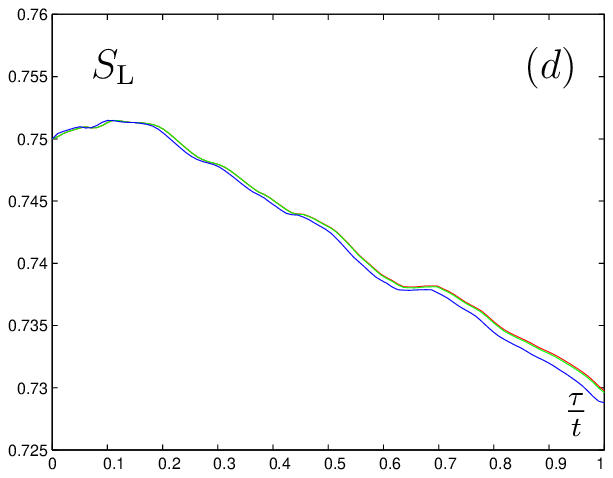}
\caption{\label{fig5}
Evolution within $0$ $\!\le$ $\!\tau$ $\!\le$ $\!t$ for an optimized cooling
with initial $p$ $\!=$ $\!0.25$ of (a) effective and (b) instantaneous controls
(red, green, and blue graphs show $x$, $y$, and $z$ component), (c) ground state
overlap of the system state, and (d) linear entropy [green, red and blue graph
show numerical integration of the Nakajima-Zwanzig equation,
time-convolutionless equation, and second order approximation of the
solution \cite{bookBreuer}, which are indistinguishable in (c)]
(system-bath coupling strength $\kappa$ $\!=$ $\!10^{-2}$,
$\omega_0$ $\!=$ $\!\frac{2\pi}{t}$, $t$ $\!=$ $\!10$, $E$ $\!=$ $\!100$).
}
\end{figure}

The time evolutions resulting from a minimization of $\Delta{S}_{\mathrm{L}}$
for the initial state Eq.~(\ref{defp}) are illustrated in Fig.~\ref{fig5}.
The $f_j$ shown in Fig.~\ref{fig5}(a) are defined by $\hat{U}(\tau)$ $\!=$
$\!\mathrm{e}^{-\frac{\mathrm{i}}{2}f_3(\tau)\hat{\sigma}_3}
\mathrm{e}^{-\frac{\mathrm{i}}{2}f_2(\tau)\hat{\sigma}_2}
\mathrm{e}^{-\frac{\mathrm{i}}{2}f_1(\tau)\hat{\sigma}_3}$,
whereas the $\omega_j$ shown in Fig.~\ref{fig5}(b) are given by
$\hat{H}_{\mathrm{S}}(\tau)$ $\!=$ $\!\sum_j\omega_j(\tau)\hat{\sigma}_j$.
The chosen constraint
$E_{}$ $\!=$ $\!\frac{1}{2}\int_{0}^{t}\!\mathrm{d}\tau
\mathrm{Tr}({H}_{\mathrm{S}}\!-\!\hat{H}_0)^2(\tau)$
can be written in terms of the $f_j$ as
$E_{}$ $\!=$ $\!\frac{1}{4}\int_{0}^{t}\mathrm{d}{t_1}\;
[\dot{f_1}^2+\dot{f_2}^2+(\dot{f_3}-\omega_0)^2
+2\dot{f_1}(\dot{f_3}-\omega_0)\cos{f_2}]$.
The overlap between the evolving system state $\hat{\varrho}(\tau)$
(in the Schr\"odinger picture) and the ground state $|0\rangle$ shown in
Fig.~\ref{fig5}(c) indicates the fast unitary system modulation through
short time population inversions without significantly altering the state purity
as verified in Fig.~\ref{fig5}(d). This can be visualized as fast
$\pi$-rotations of the state inside the Bloch sphere, which, together with
smaller rotations, here result in the final reduction of $S_{\mathrm{L}}(t)$
seen in Fig.~\ref{fig5}(d). Fig.~\ref{fig5}(d) also confirms that for the chosen
time and coupling strength, differences between various methods of
approximation are small.

\begin{figure}[ht]
\includegraphics[width=2.8cm]{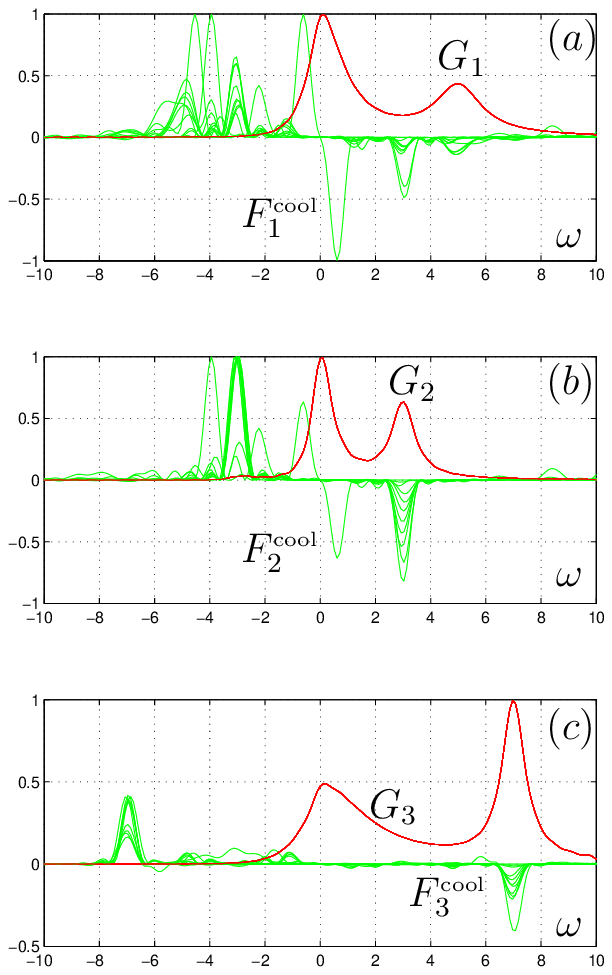}
\includegraphics[width=2.8cm]{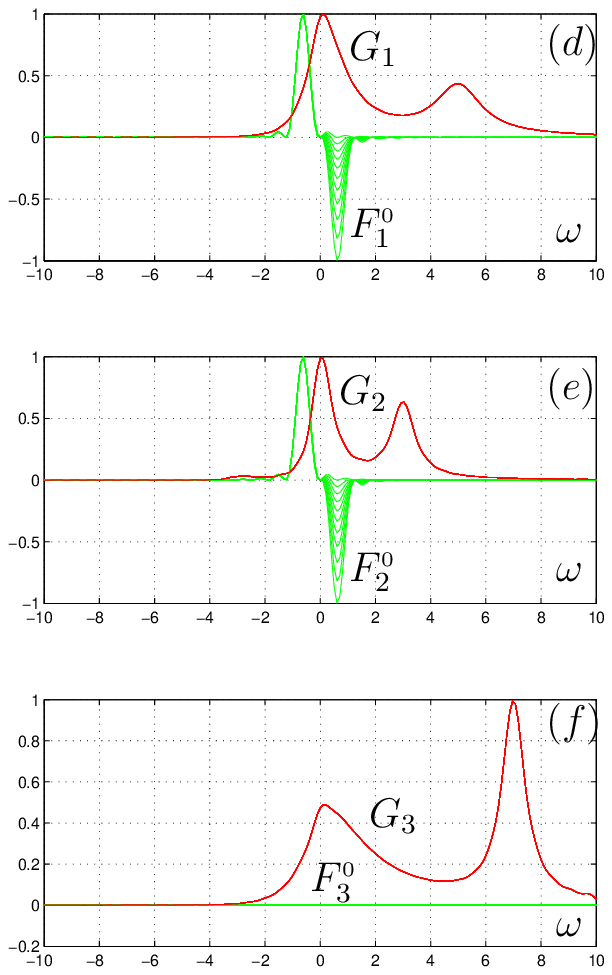}
\includegraphics[width=2.8cm]{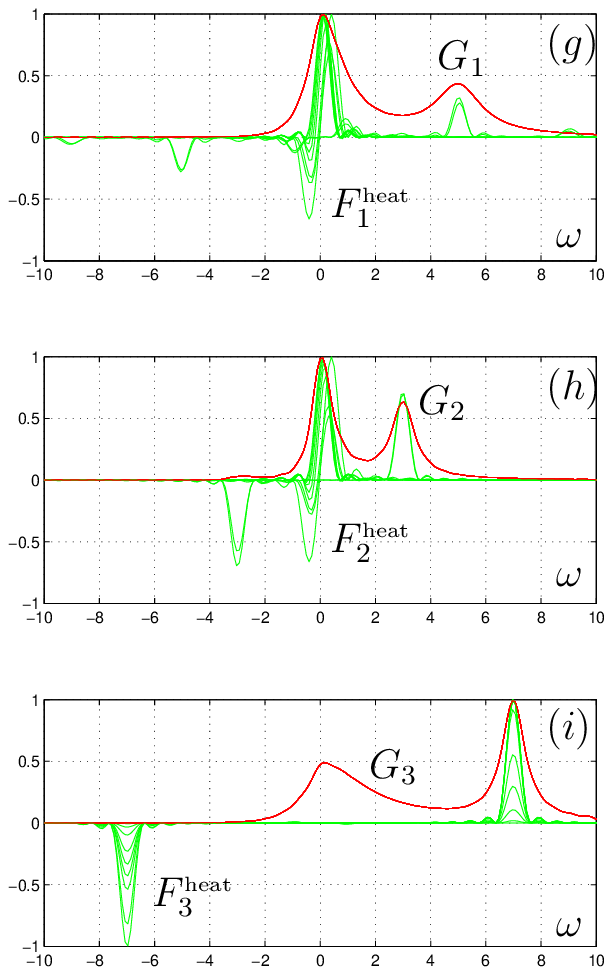}
\caption{\label{fig6}
Cooling and heating of a TLS by minimization (left: $a,b,c$) and maximization
(right: $g,h,i$) of the change of linear entropy for a given bath spectrum
(red, $G_j$) with $j=1,2,3$ denoting $x$, $y$ and $z$ component. The optimized
system spectra ($F_j$, green) are shown for an energy constraint
$E$ $\!=$ $\!10^2$ (left) and $E$ $\!=$ $\!10^3$ (right) as contrasted to the
unmodulated Hamiltonian $\hat{H}_0$
(i.e., for $E$ $\!=$ $\!0$, middle: $d,e,f$) and different initial states with
$p$ $\!=$ $\!0.001$, $0.05$, $0.1$, $\ldots$, $0.45$, $0.499$.
}
\end{figure}

In contrast to gate protection, no initial-state averaging is performed here,
i.e., Eq.~(\ref{defp}) is known. Consequently, as Fig.~\ref{fig6} shows,
the relevant components $F_{j}$ $\!\equiv$ $\!(\bm{F}_t)_{jj}(\omega)$
of the system modulation spectrum contributing to the spectral overlap
Eq.~(\ref{soverl3}) depend on the initial state $\hat{\varrho}(0)$ via
the matrix $\bm{\Gamma}$ Eq.~(\ref{Gm}). [We assume an uncorrelated
bath, i.e., $G_{jk}(\omega)$ $\!=$ $\!0$ for $j$ $\!\neq$ $\!k$ and
$G_j$ $\!\equiv$ $\!G_{jj}(\omega)$.] This influence is clearly
visible in case of a constant (unmodulated, i.e., free) Hamiltonian
(middle column), for which we set $\omega_0$ $\!=$ $\!\frac{2\pi}{t}$
with the final $t$ being in the order of the bath correlation time.
Cooling (heating) is achieved by realizing negative (positive)
$\Delta{S}_{\mathrm{L}}$ via maximum negative (positive) spectral
overlap, as shown in the left (right) column of Fig.~\ref{fig6}. This
is the opposite to system-bath decoupling, where the goal is to
minimize the overlap.

\begin{figure}[ht]
\includegraphics[width=6cm]{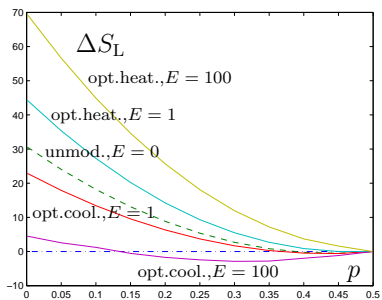}
\caption{\label{fig7}
Change of linear entropy in units of the system-bath coupling strength obtained
by minimization (attempted cooling) and maximization (attempted heating) of
$\Delta{S}_{\mathrm{L}}$ under different constraints $E$ $\!=$ $\!0,1,100$ as a
function of the initial $p$ for a bath as shown in Fig.~\ref{fig6}.
}
\end{figure}

The plots illustrate the role of the energy constraint $E$: increasing $E$
allows to establish overlap with higher frequency components of the bath
spectrum. This also suggests that for a bath spectrum with a finite frequency
cutoff, increasing $E$ beyond a certain saturation value will not lead to
further improvement of the optimization, cf. the general considerations in
App.~\ref{secA1}. In the time domain, increasing $E$ leads to more rapid changes
in the physical Hamiltonian however, requiring higher resolution of the
numerical treatment.
On the contrary, for attempted cooling (heating) by minimization (maximization)
of $\Delta{S}_{\mathrm{L}}$, a given $E$ may be too small to lead to negative
(positive) $\Delta{S}_{\mathrm{L}}$. The obtained $\Delta{S}_{\mathrm{L}}$ may
then be understood as ``reduced heating'' (``reduced cooling'') as compared to
a $\Delta{S}_{\mathrm{L}}$ obtained with an unmodulated $\hat{H}_0$. This is
shown in Fig.~\ref{fig7}.
The figure also illustrates once more that the
$\hat{\varrho}(0)$-dependence of the spectra shown in Fig.~\ref{fig6}
is accompanied by a $\hat{\varrho}(0)$-dependency of the achievable
change $\Delta{S}_{\mathrm{L}}$ for a given bath. For a maximally
mixed state in particular ($p$ $\!=$ $\!0.5$), the matrix
$\bm{\Gamma}$ Eq.~(\ref{Gm}) vanishes and with it
$\Delta{S}_{\mathrm{L}}$.

A possibility to achieve negative (positive) $\Delta{S}_{\mathrm{L}}$
by its minimization (maximization) even for weak modulation (i.e., for
small $E$), is to adapt the temperature of the bath such that for an
undriven system Hamiltonian $\hat{H}_0$, no change is observed,
$\Delta{S}_{\mathrm{L}}$ $\!=$ $\!0$, which is a necessary condition
for a system-bath equilibrium.  This is would require non-unitary
system modulation, e.g., the effect of repeated measurements
\cite{ere08,jah11,all11}.
\section{Summary and outlook}
\label{sec6}
\emph{Peculiarity of the approach}: In summary, we have considered a
way of finding a time-dependence of the system Hamiltonian over a
fixed time interval such that a given system observable attains a
desired value at the end of this interval. The peculiarity of our
approach is that it relies on knowledge of the bath coupling spectrum
and adapts the spectrum of the system modulation to it. This allows to
adjust the modulation to bandgaps or peaks in the bath coupling
spectrum. In contrast to dynamic decoupling of system and bath, which
can be achieved by shifting the entire system-modulation spectrum
beyond some assumed bath cutoff frequency, an enhancement of the
coupling requires more detailed knowledge on the peak positions of the
bath spectrum. In this way, our approach may comprise suppression and
enhancement of the system-bath coupling in a unified way for executing
more general tasks than decoherence suppression. The same approach can
also be applied to map out the bath spectrum by measuring the
coherence decay rate for a narrow-band modulation centered at
different frequencies \cite{alm11}.

As far as the controls are concerned, we here consider time-continuous
modulation of the system Hamiltonian, which allows for vastly more
freedom compared to control that is restricted to stroboscopic pulses
as in DD \cite{vio99,uhr07,kuo11}. We do not rely on rapidly changing
control fields that are required to approximate stroboscopic
$\pi$-pulses.  These features allow efficient optimization under
energy constraint. On the other hand, the generation of a sequence of
well-defined pulses may be preferable experimentally.  We may choose
the pulse timings and/or areas as continuous control-parameters and
optimize them with respect to a given bath spectrum. Hence, our
approach encompasses both pulsed and continuous modulation as special
cases.

\emph{Open issues}: An open issue of the approach is the inclusion of
higher orders in the system-bath coupling, which becomes important for
strong or resonant system-bath coupling, so that a perturbative
expansion cannot be applied.  This may be the case especially when
this coupling is to be enhanced in order to achieve a non-unitary
operation (e.g. cooling), since in this case an optimization of the
coupling may take us out of the domain of validity of the entire
approach.

Another concern regards the initial conditions. Here we have assumed a
factorized initial state of the system and bath.  This prevents us
from taking into account system-bath-interactions that may have
occurred prior to that time.  In particular, if the system is in
equilibrium with the bath, their states are entangled or correlated \cite{ere08,gorennjp12}.

An immediate problem of both higher order coupling and system-bath
correlations is that their consideration requires knowledge of the
corresponding parameters.  It may be difficult to obtain such data
with sufficient experimental precision. Moreover, its consideration
renders the theory cumbersome and the intuition gained from the
spectral overlap approach presented here is lost. A way out is offered
by replacing the ``open'' iteration loop in Fig.~\ref{fig2} with a
``closed'' loop \cite{bie09}, where the calculation of the score,
constraint and their gradients are based on actual measurements
performed on the controlled system in real time rather than on prior
model assumptions, i.e., knowledge of bath properties. Such closed
loop control would allow efficient optimization, but at the cost of
losing any insight into the physical mechanisms behind the result
obtained.

From a fundamental point of view, it is interesting to derive analytic
bounds of a desired score (under a chosen constraint) and see if this
bound can be achieved by means of some (global) optimization, i.e., if
the bound is tight.  The need for a constraint in such optimization is
not obvious if the task requires coupling enhancement, especially when
the bath spectrum has a single maximum (Appendix A).

\begin{acknowledgments}
The support of the EU (FET Open Project MIDAS), ISF and DIP is acknowledged
\end{acknowledgments}
\appendix
\section{Bound estimation}
\label{secA1}

Our goal is to give constraint-independent upper and lower bounds for the
maximum change ${P}$ $\!=$ $\!\mathrm{Tr}(\hat{P}\Delta\hat{\varrho})$
that can be achieved with a given bath and $\hat{P}$ under the condition
(\ref{condition}). We assume that $\bm{\epsilon}$, $\bm{\Gamma}$, and
$\bm{G}(\omega)$ are quadratic $(d^2\!-\!1)$-dimensional matrices.
If $\max{[\mathrm{Tr}{\bm{G}}(\omega)]}$ $\!<$ $\!\infty$, we can estimate
${P}$ by using that $\mathrm{Tr}({\bm{A}}{\bm{B}})$
$\!\le$ $\!\mathrm{Tr}({\bm{A}})\mathrm{Tr}({\bm{B}})$ for
positive semi-definite matrices ${\bm{A}}$, ${\bm{B}}$, and
applying H\"older's inequality in the form of
$\int_{-\infty}^{\infty}\mathrm{d}\omega\,|f(\omega)g(\omega)|$ $\!\le$
$\!\mathrm{sup}\{|g(\omega)|\}
\int_{-\infty}^{\infty}\mathrm{d}\omega\,|f(\omega)|$.
Decomposing ${\bm{\Gamma}}$ $\!=$ $\!{\bm{\Gamma}}_1$ $\!-$
$\!{\bm{\Gamma}}_2$ into positive semi-definite matrices
${\bm{\Gamma}}_i$ and making use of
$\frac{1}{t}\int_{-\infty}^{\infty}\mathrm{d}\omega\,
{\bm{\epsilon}_t}^\dagger(\omega){\bm{\epsilon}_t}(\omega)$
$\!=$ $\!\hat{I}$ we thus get
\begin{equation}
  -{{P}}_2\le{{P}}\le{{P}}_1,\quad
  {{P}}_i=t\,
  \sup{[\mathrm{Tr}{\bm{G}}(\omega)]}\,
  \mathrm{Tr}{\bm{\Gamma}}_i.
\end{equation}
This reveals that for given $t$ and ${\bm{G}}(\omega)$, the bounds
${{P}}_i$ depend on $\hat{\varrho}(0)$ and $\hat{P}$ via ${\bm{\Gamma}}$.
\section{\label{secA2}
         Non-commuting score}
If  Eq.~(\ref{condition}) does not hold, the following modifications must be made.
We denote by ${\bm{A}}_\pm$ $\!=$ $\!({\bm{A}}$ $\!\pm$ $\!{\bm{A}}^\dagger)/2$
the (skew) Hermitian part of a given matrix ${\bm{A}}$ $\!=$
$\!{\bm{A}}_+$ $\!+$ $\!{\bm{A}}_-$ for convenience.
Equation~(\ref{deltaP}) must be replaced with
\begin{eqnarray}
\label{A7}
  \widetilde{{P}}&=&2\mathrm{Re}
  \!\!\int_{0}^{t}\!\!\mathrm{d}t_1\!\!\int_{0}^{t_1}\!\!\mathrm{d}t_2\,
  \bigl\langle[\hat{H}_{\mathrm{I}}(t_1),\hat{P}]
  \hat{H}_{\mathrm{I}}(t_2)\bigr\rangle
  \\
  &=&{P}
  +\!\!\int_{0}^{t}\!\!\mathrm{d}t_1\!\!\int_{0}^{t_1}\!\!\mathrm{d}t_2\,
  \mathrm{Tr}\{[\hat{\varrho}_{\mathrm{tot}}(0),\hat{P}]
  \hat{H}_{\mathrm{I}}(t_2)\hat{H}_{\mathrm{I}}(t_1)\},
  \nonumber\\
\end{eqnarray}
where $\langle\cdot\rangle$ $\!=$
$\!\mathrm{Tr}[\hat{\varrho}_{\mathrm{tot}}(0)(\cdot)]$ and
${P}$ is defined in Eq.~(\ref{deltaP}). Equivalently, we can write
\begin{eqnarray}
  \widetilde{{P}}&=&
  \!\!2\!\!\int_{0}^{t}\!\!\!\mathrm{d}t_1\!\!\!\int_{0}^{t_1}\!\!\!\!
  \mathrm{d}t_2\,
  \mathrm{Tr}[\!{\bm{R}}_+(t_1,t_2){\bm{\Gamma}}_+
  \!\!+\!\!{\bm{R}}_-(t_1,t_2){\bm{\Gamma}}_-\!]\quad\quad
  \\
  &=&{P}
  -2\!\int_{0}^{t}\!\!\mathrm{d}t_1\!\!\int_{0}^{t_1}\!\!\mathrm{d}t_2\,
  \mathrm{Tr}[{\bm{R}}^\dagger(t_1,t_2){\bm{\Gamma}}_-],
\end{eqnarray}
where ${P}$ is given by Eq.~(\ref{soverl1}) together with Eq.~(\ref{dm}).
In the spectral domain, the analogous expression is
\begin{eqnarray}
  \widetilde{{P}}
  &=&2t\,\mathrm{Re}\!\int_{-\infty}^{\infty}\!\!\mathrm{d}\omega\,
  \mathrm{Tr}[{\bm{F}}_t(\omega)
  {\bm{\mathcal{G}}}(\omega)]
  \\
  &=&{P}-2t\int_{-\infty}^{\infty}\!\!\mathrm{d}\omega\,
  \mathrm{Tr}[{\bm{F}}_-(\omega){\bm{\mathcal{G}}}^\dagger(\omega)],
\end{eqnarray}
where ${P}$ is given in Eq.~(\ref{soverl3}) and
\begin{equation}
  {\bm{\mathcal{G}}}(\omega)
  =\int_{0}^{\infty}\!\mathrm{d}t\,\mathrm{e}^{\mathrm{i}\omega{t}}\,
  {\bm{\Phi}}_{}(t),
\end{equation}
is related to Eq.~(\ref{Gdef}) by
${\bm{G}}(\omega)$ $\!=$ $\!2{\bm{\mathcal{G}}}_+(\omega)$.

\begin{thebibliography}{44}
\expandafter\ifx\csname natexlab\endcsname\relax\def\natexlab#1{#1}\fi
\expandafter\ifx\csname bibnamefont\endcsname\relax
  \def\bibnamefont#1{#1}\fi
\expandafter\ifx\csname bibfnamefont\endcsname\relax
  \def\bibfnamefont#1{#1}\fi
\expandafter\ifx\csname citenamefont\endcsname\relax
  \def\citenamefont#1{#1}\fi
\expandafter\ifx\csname url\endcsname\relax
  \def\url#1{\texttt{#1}}\fi
\expandafter\ifx\csname urlprefix\endcsname\relax\def\urlprefix{URL }\fi
\providecommand{\bibinfo}[2]{#2}
\providecommand{\eprint}[2][]{\url{#2}}

\bibitem[{\citenamefont{Breuer and Petruccione}(2002)}]{bookBreuer}
\bibinfo{author}{\bibfnamefont{H.-P.} \bibnamefont{Breuer}} \bibnamefont{and}
  \bibinfo{author}{\bibfnamefont{F.}~\bibnamefont{Petruccione}},
  \emph{\bibinfo{title}{The Theory of Open Quantum Systems}}
  (\bibinfo{publisher}{Oxford University Press}, \bibinfo{address}{Oxford},
  \bibinfo{year}{2002}).

\bibitem[{\citenamefont{Bar-Gill and Kurizki}(2006)}]{BarGil06}
\bibinfo{author}{\bibfnamefont{N.}~\bibnamefont{Bar-Gill}} \bibnamefont{and}
  \bibinfo{author}{\bibfnamefont{G.}~\bibnamefont{Kurizki}},
  \bibinfo{journal}{Phys. Rev. Lett.} \textbf{\bibinfo{volume}{97}},
  \bibinfo{pages}{230402} (\bibinfo{year}{2006}).

\bibitem[{\citenamefont{Wu et~al.}(2009)\citenamefont{Wu, Kurizki, and
  Brumer}}]{WKB09}
\bibinfo{author}{\bibfnamefont{L.-A.} \bibnamefont{Wu}},
  \bibinfo{author}{\bibfnamefont{G.}~\bibnamefont{Kurizki}}, \bibnamefont{and}
  \bibinfo{author}{\bibfnamefont{P.}~\bibnamefont{Brumer}},
  \bibinfo{journal}{Phys. Rev. Lett.} \textbf{\bibinfo{volume}{102}},
  \bibinfo{pages}{080405} (\bibinfo{year}{2009}).

\bibitem[{\citenamefont{Nielsen and Chuang}(2000)}]{krausNielsen}
\bibinfo{author}{\bibfnamefont{M.~A.} \bibnamefont{Nielsen}} \bibnamefont{and}
  \bibinfo{author}{\bibfnamefont{I.~L.} \bibnamefont{Chuang}},
  \emph{\bibinfo{title}{Quantum Computation and Quantum Information}}
  (\bibinfo{publisher}{Cambridge University Press},
  \bibinfo{address}{Cambridge}, \bibinfo{year}{2000}).

\bibitem[{\citenamefont{Viola et~al.}(1999)\citenamefont{Viola, Knill, and
  Lloyd}}]{vio99}
\bibinfo{author}{\bibfnamefont{L.}~\bibnamefont{Viola}},
  \bibinfo{author}{\bibfnamefont{E.}~\bibnamefont{Knill}}, \bibnamefont{and}
  \bibinfo{author}{\bibfnamefont{S.}~\bibnamefont{Lloyd}},
  \bibinfo{journal}{Phys. Rev. Lett.} \textbf{\bibinfo{volume}{82}},
  \bibinfo{pages}{2417 } (\bibinfo{year}{1999}).

\bibitem[{\citenamefont{Uhrig}(2007)}]{uhr07}
\bibinfo{author}{\bibfnamefont{G.~S.} \bibnamefont{Uhrig}},
  \bibinfo{journal}{Phys. Rev. Lett.} \textbf{\bibinfo{volume}{98}},
  \bibinfo{pages}{100504} (\bibinfo{year}{2007}).

\bibitem[{\citenamefont{Kuo and Lidar}(2011)}]{kuo11}
\bibinfo{author}{\bibfnamefont{W.-J.} \bibnamefont{Kuo}} \bibnamefont{and}
  \bibinfo{author}{\bibfnamefont{D.~A.} \bibnamefont{Lidar}}
  (\bibinfo{year}{2011}), \bibinfo{note}{http://arxiv.org/abs/1106.2151}.

\bibitem[{\citenamefont{Cai et~al.}(2011)\citenamefont{Cai, Jelezko, Plenio,
  and Retzker}}]{cai11}
\bibinfo{author}{\bibfnamefont{J.-M.} \bibnamefont{Cai}},
  \bibinfo{author}{\bibfnamefont{F.}~\bibnamefont{Jelezko}},
  \bibinfo{author}{\bibfnamefont{M.~B.} \bibnamefont{Plenio}},
  \bibnamefont{and} \bibinfo{author}{\bibfnamefont{A.}~\bibnamefont{Retzker}}
  (\bibinfo{year}{2011}), \bibinfo{note}{http://arxiv.org/abs/1111.0930}.

\bibitem[{\citenamefont{Kofman and Kurizki}(2001)}]{kof01}
\bibinfo{author}{\bibfnamefont{A.~G.} \bibnamefont{Kofman}} \bibnamefont{and}
  \bibinfo{author}{\bibfnamefont{G.}~\bibnamefont{Kurizki}},
  \bibinfo{journal}{Phys. Rev. Lett.} \textbf{\bibinfo{volume}{87}},
  \bibinfo{pages}{270405} (\bibinfo{year}{2001}).

\bibitem[{\citenamefont{Barone et~al.}(2004)\citenamefont{Barone, Kurizki, and
  Kofman}}]{Barone04}
\bibinfo{author}{\bibfnamefont{A.}~\bibnamefont{Barone}},
  \bibinfo{author}{\bibfnamefont{G.}~\bibnamefont{Kurizki}}, \bibnamefont{and}
  \bibinfo{author}{\bibfnamefont{A.~G.} \bibnamefont{Kofman}},
  \bibinfo{journal}{Phys. Rev. Lett.} \textbf{\bibinfo{volume}{92}},
  \bibinfo{pages}{200403} (\bibinfo{year}{2004}).

\bibitem[{\citenamefont{Kofman and Kurizki}(2004)}]{kof04}
\bibinfo{author}{\bibfnamefont{A.~G.} \bibnamefont{Kofman}} \bibnamefont{and}
  \bibinfo{author}{\bibfnamefont{G.}~\bibnamefont{Kurizki}},
  \bibinfo{journal}{Phys. Rev. Lett.} \textbf{\bibinfo{volume}{93}},
  \bibinfo{pages}{130406} (\bibinfo{year}{2004}).

\bibitem[{\citenamefont{Gordon et~al.}(2007)\citenamefont{Gordon, Erez, and
  Kurizki}}]{JPB}
\bibinfo{author}{\bibfnamefont{G.}~\bibnamefont{Gordon}},
  \bibinfo{author}{\bibfnamefont{N.}~\bibnamefont{Erez}}, \bibnamefont{and}
  \bibinfo{author}{\bibfnamefont{G.}~\bibnamefont{Kurizki}},
  \bibinfo{journal}{J. Phys. B: At. Mol. Opt. Phys.}
  \textbf{\bibinfo{volume}{40}}, \bibinfo{pages}{S75} (\bibinfo{year}{2007}).

\bibitem[{\citenamefont{Gordon et~al.}(2008)\citenamefont{Gordon, Kurizki, and
  Lidar}}]{gkl08}
\bibinfo{author}{\bibfnamefont{G.}~\bibnamefont{Gordon}},
  \bibinfo{author}{\bibfnamefont{G.}~\bibnamefont{Kurizki}}, \bibnamefont{and}
  \bibinfo{author}{\bibfnamefont{D.~A.} \bibnamefont{Lidar}},
  \bibinfo{journal}{Phys. Rev. Lett.} \textbf{\bibinfo{volume}{101}},
  \bibinfo{pages}{010403} (\bibinfo{year}{2008}).

\bibitem[{\citenamefont{Alicki}(1979)}]{Alicki79}
\bibinfo{author}{\bibfnamefont{R.}~\bibnamefont{Alicki}}, \bibinfo{journal}{J.
  Phys. A} \textbf{\bibinfo{volume}{12}}, \bibinfo{pages}{L103}
  (\bibinfo{year}{1979}).

\bibitem[{\citenamefont{Lindblad}(1983)}]{Lindbladbook}
\bibinfo{author}{\bibfnamefont{G.}~\bibnamefont{Lindblad}},
  \emph{\bibinfo{title}{Non-Equilibrium Entropy and Irreversibility}}
  (\bibinfo{publisher}{D. Reidel}, \bibinfo{address}{Holland},
  \bibinfo{year}{1983});
\bibinfo{author}{\bibfnamefont{J.}~\bibnamefont{Gemmer}},
  \bibinfo{author}{\bibfnamefont{M.}~\bibnamefont{Michel}}, \bibnamefont{and}
  \bibinfo{author}{\bibfnamefont{G.}~\bibnamefont{Mahler}},
  \emph{\bibinfo{title}{Quantum Thermodynamics: Emergence of Thermodynamic
  Behavior Within Composite Quantum Systems}}, vol. \bibinfo{volume}{657} of
  \emph{\bibinfo{series}{Lecture Notes in Physics}}
  (\bibinfo{publisher}{Springer}, \bibinfo{address}{Berlin Heidelberg},
  \bibinfo{year}{2004});
\bibinfo{author}{\bibfnamefont{R.}~\bibnamefont{Alicki}},
  \bibinfo{author}{\bibfnamefont{M.}~\bibnamefont{Horodecki}},
  \bibinfo{author}{\bibfnamefont{P.}~\bibnamefont{Horodecki}},
  \bibnamefont{and}
  \bibinfo{author}{\bibfnamefont{R.}~\bibnamefont{Horodecki}},
  \bibinfo{journal}{Open Systems amp; Information Dynamics}
  \textbf{\bibinfo{volume}{11}}, \bibinfo{pages}{205} (\bibinfo{year}{2004}),
  ISSN \bibinfo{issn}{1230-1612},
  \bibinfo{note}{10.1023/B:OPSY.0000047566.72717.71};
\bibinfo{author}{\bibfnamefont{H.}~\bibnamefont{Spohn}}, \bibinfo{journal}{J.
  Math. Phys.} \textbf{\bibinfo{volume}{19}}, \bibinfo{pages}{1227}
  (\bibinfo{year}{1978});
\bibinfo{author}{\bibfnamefont{J.}~\bibnamefont{Gemmer}},
  \bibinfo{author}{\bibfnamefont{A.}~\bibnamefont{Otte}}, \bibnamefont{and}
  \bibinfo{author}{\bibfnamefont{G.}~\bibnamefont{Mahler}},
  \bibinfo{journal}{Phys. Rev. Lett.} \textbf{\bibinfo{volume}{86}},
  \bibinfo{pages}{1927} (\bibinfo{year}{2001}).

\bibitem[{\citenamefont{Gordon et~al.}(2009)\citenamefont{Gordon, Bensky,
  Gelbwaser-Klimovsky, Rao, Erez, and Kurizki}}]{gor09}
\bibinfo{author}{\bibfnamefont{G.}~\bibnamefont{Gordon}},
  \bibinfo{author}{\bibfnamefont{G.}~\bibnamefont{Bensky}},
  \bibinfo{author}{\bibfnamefont{D.}~\bibnamefont{Gelbwaser-Klimovsky}},
  \bibinfo{author}{\bibfnamefont{D.}~\bibnamefont{Rao}},
  \bibinfo{author}{\bibfnamefont{N.}~\bibnamefont{Erez}}, \bibnamefont{and}
  \bibinfo{author}{\bibfnamefont{G.}~\bibnamefont{Kurizki}},
  \bibinfo{journal}{New J. Phys.} \textbf{\bibinfo{volume}{11}},
  \bibinfo{pages}{123025} (\bibinfo{year}{2009}).

\bibitem[{\citenamefont{Gordon et~al.}(2010)\citenamefont{Gordon, Rao, and
  Kurizki}}]{gorennjp12}
\bibinfo{author}{\bibfnamefont{G.}~\bibnamefont{Gordon}},
  \bibinfo{author}{\bibfnamefont{D.~D.~B.} \bibnamefont{Rao}},
  \bibnamefont{and} \bibinfo{author}{\bibfnamefont{G.}~\bibnamefont{Kurizki}},
  \bibinfo{journal}{New Journal of Physics} \textbf{\bibinfo{volume}{12}},
  \bibinfo{pages}{053033} (\bibinfo{year}{2010}).

\bibitem[{\citenamefont{\'Alvarez et~al.}(2010)\citenamefont{\'Alvarez, Rao,
  Frydman, and Kurizki}}]{Gonzalo10}
\bibinfo{author}{\bibfnamefont{G.~A.} \bibnamefont{\'Alvarez}},
  \bibinfo{author}{\bibfnamefont{D.~D.~B.} \bibnamefont{Rao}},
  \bibinfo{author}{\bibfnamefont{L.}~\bibnamefont{Frydman}}, \bibnamefont{and}
  \bibinfo{author}{\bibfnamefont{G.}~\bibnamefont{Kurizki}},
  \bibinfo{journal}{Phys. Rev. Lett.} \textbf{\bibinfo{volume}{105}},
  \bibinfo{pages}{160401} (\bibinfo{year}{2010}).

\bibitem[{\citenamefont{Chan et~al.}(2011)\citenamefont{Chan, Alegre,
  Safavi-Naeini, Hill, Krause, Gr{\"o}blacher, Aspelmeyer, and
  Painter}}]{cha11}
\bibinfo{author}{\bibfnamefont{J.}~\bibnamefont{Chan}},
  \bibinfo{author}{\bibfnamefont{T.~P.~M.} \bibnamefont{Alegre}},
  \bibinfo{author}{\bibfnamefont{A.~H.} \bibnamefont{Safavi-Naeini}},
  \bibinfo{author}{\bibfnamefont{J.~T.} \bibnamefont{Hill}},
  \bibinfo{author}{\bibfnamefont{A.}~\bibnamefont{Krause}},
  \bibinfo{author}{\bibfnamefont{S.}~\bibnamefont{Gr{\"o}blacher}},
  \bibinfo{author}{\bibfnamefont{M.}~\bibnamefont{Aspelmeyer}},
  \bibnamefont{and} \bibinfo{author}{\bibfnamefont{O.}~\bibnamefont{Painter}},
  \bibinfo{journal}{Nature} \textbf{\bibinfo{volume}{478}}, \bibinfo{pages}{89}
  (\bibinfo{year}{2011}).

\bibitem[{\citenamefont{Harel et~al.}(1996)\citenamefont{Harel, Kurizki,
  McIver, and Coutsias}}]{Harel96}
\bibinfo{author}{\bibfnamefont{G.}~\bibnamefont{Harel}},
  \bibinfo{author}{\bibfnamefont{G.}~\bibnamefont{Kurizki}},
  \bibinfo{author}{\bibfnamefont{J.~K.} \bibnamefont{McIver}},
  \bibnamefont{and} \bibinfo{author}{\bibfnamefont{E.}~\bibnamefont{Coutsias}},
  \bibinfo{journal}{Phys. Rev. A} \textbf{\bibinfo{volume}{53}},
  \bibinfo{pages}{4534} (\bibinfo{year}{1996});
\bibinfo{author}{\bibfnamefont{B.~M.} \bibnamefont{Garraway}},
  \bibinfo{author}{\bibfnamefont{B.}~\bibnamefont{Sherman}},
  \bibinfo{author}{\bibfnamefont{H.}~\bibnamefont{Moya-Cessa}},
  \bibinfo{author}{\bibfnamefont{P.~L.} \bibnamefont{Knight}},
  \bibnamefont{and} \bibinfo{author}{\bibfnamefont{G.}~\bibnamefont{Kurizki}},
  \bibinfo{journal}{Phys. Rev. A} \textbf{\bibinfo{volume}{49}},
  \bibinfo{pages}{535} (\bibinfo{year}{1994}).

\bibitem[{\citenamefont{Kofman and Kurizki}(2000)}]{kof00}
\bibinfo{author}{\bibfnamefont{A.~G.} \bibnamefont{Kofman}} \bibnamefont{and}
  \bibinfo{author}{\bibfnamefont{G.}~\bibnamefont{Kurizki}},
  \bibinfo{journal}{Nature (London)} \textbf{\bibinfo{volume}{405}},
  \bibinfo{pages}{546} (\bibinfo{year}{2000}).

\bibitem[{\citenamefont{Opatrn\'y et~al.}(2000)\citenamefont{Opatrn\'y,
  Kurizki, and Welsch}}]{Opatrny00}
\bibinfo{author}{\bibfnamefont{T.}~\bibnamefont{Opatrn\'y}},
  \bibinfo{author}{\bibfnamefont{G.}~\bibnamefont{Kurizki}}, \bibnamefont{and}
  \bibinfo{author}{\bibfnamefont{D.-G.} \bibnamefont{Welsch}},
  \bibinfo{journal}{Phys. Rev. A} \textbf{\bibinfo{volume}{61}},
  \bibinfo{pages}{032302} (\bibinfo{year}{2000}).

\bibitem[{\citenamefont{Scully et~al.}(2003)\citenamefont{Scully, Zubairy,
  Agarwal, and Walther}}]{Scully03}
\bibinfo{author}{\bibfnamefont{M.~O.} \bibnamefont{Scully}},
  \bibinfo{author}{\bibfnamefont{M.~S.} \bibnamefont{Zubairy}},
  \bibinfo{author}{\bibfnamefont{G.~S.} \bibnamefont{Agarwal}},
  \bibnamefont{and} \bibinfo{author}{\bibfnamefont{H.}~\bibnamefont{Walther}},
  \bibinfo{journal}{Science} \textbf{\bibinfo{volume}{299}},
  \bibinfo{pages}{862} (\bibinfo{year}{2003});
\bibinfo{author}{\bibfnamefont{M.~O.} \bibnamefont{Scully}},
  \bibinfo{journal}{Phys. Rev. Lett.} \textbf{\bibinfo{volume}{88}},
  \bibinfo{pages}{050602} (\bibinfo{year}{2002});
\bibinfo{author}{\bibfnamefont{M.~O.} \bibnamefont{Scully}},
  \bibinfo{journal}{Phys. Rev. Lett.} \textbf{\bibinfo{volume}{104}},
  \bibinfo{pages}{207701} (\bibinfo{year}{2010});
\bibinfo{author}{\bibfnamefont{M.~O.} \bibnamefont{Scully}},
  \bibinfo{author}{\bibfnamefont{K.~R.} \bibnamefont{Chapin}},
  \bibinfo{author}{\bibfnamefont{K.~E.} \bibnamefont{Dorfman}},
  \bibinfo{author}{\bibfnamefont{M.~B.} \bibnamefont{Kim}}, \bibnamefont{and}
  \bibinfo{author}{\bibfnamefont{A.}~\bibnamefont{Svidzinsky}},
  \bibinfo{journal}{Proceedings of the National Academy of Sciences}
  (\bibinfo{year}{2011}),
  \eprint{http://www.pnas.org/content/early/2011/08/25/ 1110234108.full.pdf+htm%
l};
\bibinfo{author}{\bibfnamefont{T.}~\bibnamefont{Opatrn\'{y}}},
  \bibinfo{journal}{American Journal of Physics} \textbf{\bibinfo{volume}{73}},
  \bibinfo{pages}{63} (\bibinfo{year}{2005}).

\bibitem[{\citenamefont{van Grondelle and Novoderezhkin}(2010)}]{gro10}
\bibinfo{author}{\bibfnamefont{R.}~\bibnamefont{van Grondelle}}
  \bibnamefont{and} \bibinfo{author}{\bibfnamefont{V.~I.}
  \bibnamefont{Novoderezhkin}}, \bibinfo{journal}{Nature}
  \textbf{\bibinfo{volume}{463}}, \bibinfo{pages}{614} (\bibinfo{year}{2010}).

\bibitem[{\citenamefont{Scholes}(2011)}]{scho11}
\bibinfo{author}{\bibfnamefont{G.~D.} \bibnamefont{Scholes}},
  \bibinfo{journal}{Nature Physics} \textbf{\bibinfo{volume}{7}},
  \bibinfo{pages}{448} (\bibinfo{year}{2011}).

\bibitem[{\citenamefont{Schaller and Klimov}(2004)}]{sch04}
\bibinfo{author}{\bibfnamefont{R.~D.} \bibnamefont{Schaller}} \bibnamefont{and}
  \bibinfo{author}{\bibfnamefont{V.~I.} \bibnamefont{Klimov}},
  \bibinfo{journal}{Phys. Rev. Lett.} \textbf{\bibinfo{volume}{92}},
  \bibinfo{pages}{186601} (\bibinfo{year}{2004});
\bibinfo{author}{\bibfnamefont{R.~D.} \bibnamefont{Schaller}},
  \bibinfo{author}{\bibfnamefont{M.}~\bibnamefont{Sykora}},
  \bibinfo{author}{\bibfnamefont{J.~M.} \bibnamefont{Pietryga}},
  \bibnamefont{and} \bibinfo{author}{\bibfnamefont{V.~I.}
  \bibnamefont{Klimov}}, \bibinfo{journal}{Nano Lett.}
  \textbf{\bibinfo{volume}{6}}, \bibinfo{pages}{424} (\bibinfo{year}{2006}).

\bibitem[{\citenamefont{Blankenship et~al.}(2011)\citenamefont{Blankenship,
  Tiede, Barber, Brudvig, Fleming, Ghirardi, Gunner, Junge, Kramer, Melis
  et~al.}}]{bla11}
\bibinfo{author}{\bibfnamefont{R.~E.} \bibnamefont{Blankenship}},
  \bibinfo{author}{\bibfnamefont{D.~M.} \bibnamefont{Tiede}},
  \bibinfo{author}{\bibfnamefont{J.}~\bibnamefont{Barber}},
  \bibinfo{author}{\bibfnamefont{G.~W.} \bibnamefont{Brudvig}},
  \bibinfo{author}{\bibfnamefont{G.}~\bibnamefont{Fleming}},
  \bibinfo{author}{\bibfnamefont{M.}~\bibnamefont{Ghirardi}},
  \bibinfo{author}{\bibfnamefont{M.~R.} \bibnamefont{Gunner}},
  \bibinfo{author}{\bibfnamefont{W.}~\bibnamefont{Junge}},
  \bibinfo{author}{\bibfnamefont{D.~M.} \bibnamefont{Kramer}},
  \bibinfo{author}{\bibfnamefont{A.}~\bibnamefont{Melis}},
  \bibnamefont{et~al.}, \bibinfo{journal}{Science}
  \textbf{\bibinfo{volume}{332}}, \bibinfo{pages}{805} (\bibinfo{year}{2011}).

\bibitem[{\citenamefont{Escher et~al.}(2011)\citenamefont{Escher, Bensky,
  Clausen, and Kurizki}}]{esc11}
\bibinfo{author}{\bibfnamefont{B.~M.} \bibnamefont{Escher}},
  \bibinfo{author}{\bibfnamefont{G.}~\bibnamefont{Bensky}},
  \bibinfo{author}{\bibfnamefont{J.}~\bibnamefont{Clausen}}, \bibnamefont{and}
  \bibinfo{author}{\bibfnamefont{G.}~\bibnamefont{Kurizki}},
  \bibinfo{journal}{J. Phys. B: At. Mol. Opt. Phys.}
  \textbf{\bibinfo{volume}{44}}, \bibinfo{pages}{154015}
  (\bibinfo{year}{2011}).

\bibitem[{\citenamefont{Vollbrecht et~al.}(2011)\citenamefont{Vollbrecht,
  Muschik, and Cirac}}]{vol11}
\bibinfo{author}{\bibfnamefont{K.~G.~H.} \bibnamefont{Vollbrecht}},
  \bibinfo{author}{\bibfnamefont{C.~A.} \bibnamefont{Muschik}},
  \bibnamefont{and} \bibinfo{author}{\bibfnamefont{J.~I.} \bibnamefont{Cirac}},
  \bibinfo{journal}{Phys. Rev. Lett.} \textbf{\bibinfo{volume}{107}},
  \bibinfo{pages}{120502} (\bibinfo{year}{2011}).

\bibitem[{\citenamefont{Bhaktavatsala~Rao
  et~al.}(2011)\citenamefont{Bhaktavatsala~Rao, Bar-Gill, and
  Kurizki}}]{durga11}
\bibinfo{author}{\bibfnamefont{D.~D.} \bibnamefont{Bhaktavatsala~Rao}},
  \bibinfo{author}{\bibfnamefont{N.}~\bibnamefont{Bar-Gill}}, \bibnamefont{and}
  \bibinfo{author}{\bibfnamefont{G.}~\bibnamefont{Kurizki}},
  \bibinfo{journal}{Phys. Rev. Lett.} \textbf{\bibinfo{volume}{106}},
  \bibinfo{pages}{010404} (\bibinfo{year}{2011});
\bibinfo{author}{\bibfnamefont{D.}~\bibnamefont{Braun}},
  \bibinfo{journal}{Phys. Rev. Lett.} \textbf{\bibinfo{volume}{89}},
  \bibinfo{pages}{277901} (\bibinfo{year}{2002});
\bibinfo{author}{\bibfnamefont{G.}~\bibnamefont{Kurizki}},
  \bibinfo{journal}{Phys. Rev. A} \textbf{\bibinfo{volume}{42}},
  \bibinfo{pages}{2915} (\bibinfo{year}{1990});
\bibinfo{author}{\bibfnamefont{G.}~\bibnamefont{Kurizki}},
  \bibinfo{author}{\bibfnamefont{A.~G.} \bibnamefont{Kofman}},
  \bibnamefont{and} \bibinfo{author}{\bibfnamefont{V.}~\bibnamefont{Yudson}},
  \bibinfo{journal}{Phys. Rev. A} \textbf{\bibinfo{volume}{53}},
  \bibinfo{pages}{R35} (\bibinfo{year}{1996}).

\bibitem[{\citenamefont{Gordon and Kurizki}(2006)}]{gor06b}
\bibinfo{author}{\bibfnamefont{G.}~\bibnamefont{Gordon}} \bibnamefont{and}
  \bibinfo{author}{\bibfnamefont{G.}~\bibnamefont{Kurizki}},
  \bibinfo{journal}{Phys. Rev. Lett.} \textbf{\bibinfo{volume}{97}},
  \bibinfo{pages}{110503} (\bibinfo{year}{2006}).

\bibitem[{\citenamefont{Cai et~al.}(2010)\citenamefont{Cai, Guerreschi, and
  Briegel}}]{cai10}
\bibinfo{author}{\bibfnamefont{J.}~\bibnamefont{Cai}},
  \bibinfo{author}{\bibfnamefont{G.~G.} \bibnamefont{Guerreschi}},
  \bibnamefont{and} \bibinfo{author}{\bibfnamefont{H.~J.}
  \bibnamefont{Briegel}}, \bibinfo{journal}{Phys. Rev. Lett.}
  \textbf{\bibinfo{volume}{104}}, \bibinfo{pages}{220502}
  (\bibinfo{year}{2010}).

\bibitem[{\citenamefont{Scholak et~al.}(2011)\citenamefont{Scholak, de~Melo,
  Wellens, Mintert, and Buchleitner}}]{sch11}
\bibinfo{author}{\bibfnamefont{T.}~\bibnamefont{Scholak}},
  \bibinfo{author}{\bibfnamefont{F.}~\bibnamefont{de~Melo}},
  \bibinfo{author}{\bibfnamefont{T.}~\bibnamefont{Wellens}},
  \bibinfo{author}{\bibfnamefont{F.}~\bibnamefont{Mintert}}, \bibnamefont{and}
  \bibinfo{author}{\bibfnamefont{A.}~\bibnamefont{Buchleitner}},
  \bibinfo{journal}{Phys. Rev. E} \textbf{\bibinfo{volume}{83}},
  \bibinfo{pages}{021912} (\bibinfo{year}{2011}).

\bibitem[{\citenamefont{Pechen and Tannor}(2011)}]{pec11}
\bibinfo{author}{\bibfnamefont{A.~N.} \bibnamefont{Pechen}} \bibnamefont{and}
  \bibinfo{author}{\bibfnamefont{D.~J.} \bibnamefont{Tannor}},
  \bibinfo{journal}{Phys. Rev. Lett.} \textbf{\bibinfo{volume}{106}},
  \bibinfo{pages}{120402} (\bibinfo{year}{2011}).

\bibitem[{\citenamefont{Kofman and Kurizki}(2005)}]{kof05}
\bibinfo{author}{\bibfnamefont{A.~G.} \bibnamefont{Kofman}} \bibnamefont{and}
  \bibinfo{author}{\bibfnamefont{G.}~\bibnamefont{Kurizki}},
  \bibinfo{journal}{IEEE Trans. Nanotechnology} \textbf{\bibinfo{volume}{4}},
  \bibinfo{pages}{116} (\bibinfo{year}{2005}).

\bibitem[{\citenamefont{Kofman and Kurizki}(1996)}]{kof96}
\bibinfo{author}{\bibfnamefont{A.~G.} \bibnamefont{Kofman}} \bibnamefont{and}
  \bibinfo{author}{\bibfnamefont{G.}~\bibnamefont{Kurizki}},
  \bibinfo{journal}{Phys. Rev. A} \textbf{\bibinfo{volume}{54}},
  \bibinfo{pages}{R3750} (\bibinfo{year}{1996}).

\bibitem[{\citenamefont{Clausen et~al.}(2010)\citenamefont{Clausen, Bensky, and
  Kurizki}}]{cla10}
\bibinfo{author}{\bibfnamefont{J.}~\bibnamefont{Clausen}},
  \bibinfo{author}{\bibfnamefont{G.}~\bibnamefont{Bensky}}, \bibnamefont{and}
  \bibinfo{author}{\bibfnamefont{G.}~\bibnamefont{Kurizki}},
  \bibinfo{journal}{Phys. Rev. Lett.} \textbf{\bibinfo{volume}{104}},
  \bibinfo{pages}{040401} (\bibinfo{year}{2010}).

\bibitem[{\citenamefont{Dankert}(2005)}]{dan05}
\bibinfo{author}{\bibfnamefont{C.}~\bibnamefont{Dankert}}, Master's thesis,
  \bibinfo{school}{University of Waterloo}, \bibinfo{address}{Ontario, Canada}
  (\bibinfo{year}{2005}).

\bibitem[{\citenamefont{Storcz et~al.}(2005)\citenamefont{Storcz, Hartmann,
  Kohler, and Wilhelm}}]{sto05}
\bibinfo{author}{\bibfnamefont{M.~J.} \bibnamefont{Storcz}},
  \bibinfo{author}{\bibfnamefont{U.}~\bibnamefont{Hartmann}},
  \bibinfo{author}{\bibfnamefont{S.}~\bibnamefont{Kohler}}, \bibnamefont{and}
  \bibinfo{author}{\bibfnamefont{F.~K.} \bibnamefont{Wilhelm}},
  \bibinfo{journal}{Phys. Rev. B} \textbf{\bibinfo{volume}{72}},
  \bibinfo{pages}{235321} (\bibinfo{year}{2005}).

\bibitem[{\citenamefont{Erez et~al.}(2008)\citenamefont{Erez, Gordon, Nest, and
  Kurizki}}]{ere08}
\bibinfo{author}{\bibfnamefont{N.}~\bibnamefont{Erez}},
  \bibinfo{author}{\bibfnamefont{G.}~\bibnamefont{Gordon}},
  \bibinfo{author}{\bibfnamefont{M.}~\bibnamefont{Nest}}, \bibnamefont{and}
  \bibinfo{author}{\bibfnamefont{G.}~\bibnamefont{Kurizki}},
  \bibinfo{journal}{Nature} \textbf{\bibinfo{volume}{452}},
  \bibinfo{pages}{724} (\bibinfo{year}{2008}).

\bibitem[{\citenamefont{Jahnke and Mahler}(2011)}]{jah11}
\bibinfo{author}{\bibfnamefont{T.}~\bibnamefont{Jahnke}} \bibnamefont{and}
  \bibinfo{author}{\bibfnamefont{G.}~\bibnamefont{Mahler}},
  \bibinfo{journal}{Phys. Rev. E} \textbf{\bibinfo{volume}{84}},
  \bibinfo{pages}{011129} (\bibinfo{year}{2011}).

\bibitem[{\citenamefont{Allahverdyan et~al.}(2011)\citenamefont{Allahverdyan,
  Hovhannisyan, Janzing, and Mahler}}]{all11}
\bibinfo{author}{\bibfnamefont{A.~E.} \bibnamefont{Allahverdyan}},
  \bibinfo{author}{\bibfnamefont{K.~V.} \bibnamefont{Hovhannisyan}},
  \bibinfo{author}{\bibfnamefont{D.}~\bibnamefont{Janzing}}, \bibnamefont{and}
  \bibinfo{author}{\bibfnamefont{G.}~\bibnamefont{Mahler}},
  \bibinfo{journal}{Phys. Rev. E} \textbf{\bibinfo{volume}{84}},
  \bibinfo{pages}{041109} (\bibinfo{year}{2011}).

\bibitem[{\citenamefont{Almog et~al.}(2011)\citenamefont{Almog, Sagi, Gordon,
  Bensky, Kurizki, and Davidson}}]{alm11}
\bibinfo{author}{\bibfnamefont{I.}~\bibnamefont{Almog}},
  \bibinfo{author}{\bibfnamefont{Y.}~\bibnamefont{Sagi}},
  \bibinfo{author}{\bibfnamefont{G.}~\bibnamefont{Gordon}},
  \bibinfo{author}{\bibfnamefont{G.}~\bibnamefont{Bensky}},
  \bibinfo{author}{\bibfnamefont{G.}~\bibnamefont{Kurizki}}, \bibnamefont{and}
  \bibinfo{author}{\bibfnamefont{N.}~\bibnamefont{Davidson}},
  \bibinfo{journal}{J. Phys. B: At. Mol. Opt. Phys.}
  \textbf{\bibinfo{volume}{44}}, \bibinfo{pages}{154006}
  (\bibinfo{year}{2011}).

\bibitem[{\citenamefont{Biercuk et~al.}(2009)\citenamefont{Biercuk, Uys,
  VanDevender, Shiga, Itano, and Bollinger}}]{bie09}
\bibinfo{author}{\bibfnamefont{M.~J.} \bibnamefont{Biercuk}},
  \bibinfo{author}{\bibfnamefont{H.}~\bibnamefont{Uys}},
  \bibinfo{author}{\bibfnamefont{A.~P.} \bibnamefont{VanDevender}},
  \bibinfo{author}{\bibfnamefont{N.}~\bibnamefont{Shiga}},
  \bibinfo{author}{\bibfnamefont{W.~M.} \bibnamefont{Itano}}, \bibnamefont{and}
  \bibinfo{author}{\bibfnamefont{J.~J.} \bibnamefont{Bollinger}},
  \bibinfo{journal}{Nature} \textbf{\bibinfo{volume}{458}},
  \bibinfo{pages}{996} (\bibinfo{year}{2009}).

\end{thebibliography}

\end{document}